# The pathway to chirality in elemental tellurium

Yuxing Zhou[1], Stephen R. Elliott[2], Daniel F. Thomas du Toit[1], Wei Zhang[3]*, Volker L. Deringer[1]*

[1] Inorganic Chemistry Laboratory, Department of Chemistry, University of Oxford, Oxford OX1 3QR, United Kingdom

[2] Physical and Theoretical Chemistry Laboratory, Department of Chemistry, University of Oxford, Oxford OX1 3QZ, United Kingdom

[3] Center for Alloy Innovation and Design (CAID), State Key Laboratory for Mechanical Behavior of Materials, Xi'an Jiaotong University, Xi'an 710049, China

* Corresponding authors. Email: wzhang0@mail.xjtu.edu.cn; volker.deringer@chem.ox.ac.uk

**Abstract:** Chiral crystals, like chiral molecules, cannot be superimposed onto their mirror images—a fundamental property that has been linked to interesting physical behavior and exploited in functional devices. Among the simplest inorganic systems with crystallographic chirality, elemental tellurium adopts crystal structures with right- or left-handed chains. However, understanding the formation mechanisms of those structures has been difficult due to the rapid crystallization of Te, which reaches the spatial and temporal resolution limits of even the most advanced experiments. Here, we report ultra-large-scale, quantum-mechanically accurate simulations that reveal mechanisms of crystallization and the origin of crystallographic chirality in solid Te. We identify a characteristic, disordered cube-like structural motif—a transient bonding environment with only nanosecond lifetime—that enables both the rapid crystallization of Te and mediates chirality transfer. Based on the resulting microscopic understanding, we are able to explain the switching behavior of Te-based electrical devices.

Chirality, or handedness, is found in many organic molecules and also in inorganic solids: e.g., elemental selenium and tellurium, the common quartz polymorph of silica (α-SiO$_2$), and cinnabar (α-HgS), which all crystallize in the same chiral space groups. The ambient form of tellurium, α-Te, is among the simplest inorganic systems for studying chirality (*1*). Two types of chirality are observed in α-Te, at different length scales: morphographic and crystallographic chirality. Morphographic chirality refers to the particle shape and is present in a wide range of organic and biological systems (*2–4*) as well as in inorganic materials (*5–8*)—indeed, its formation in Te nanoparticles was recently suggested to be driven by screw dislocations during crystallization, supported by atomic-resolution images from scanning transmission electron microscopy (STEM) (*9*). Crystallographic chirality arises from the helical arrangement of atoms in the $P3_121$ (right-handed) or $P3_221$ (left-handed) structure, which is of interest because of its direct connection with physical properties. An example is the interplay of crystallographic chirality and electron spins in α-Te (*10*), which allows for the generation of spin polarization (*11*) and has been used in emerging spintronic devices (*12, 13*). In addition, low-dimensional Te nanostructures (e.g., nano-tubes, -flakes, and -wires) have been used in functional devices (*14, 15*). To obtain the desired morphography, chemical synthesis of these nanostructures can be controlled via nucleation and growth under different conditions (*16–18*).

Te has a very strong crystallization tendency: the liquid element will quickly, even explosively, crystallize into helical structures at ambient temperature (*18*) or below (*19*). Such phase transition is accompanied by a large change in electrical resistance from a metallic liquid state to a semiconducting crystalline state (*20*). These features make Te a promising candidate for selector devices (*21*). The selector is a critical component for high-density non-volatile memory products, which suppresses the leakage current in three-dimensionally stacked cross-point arrays (*22–24*). For electrical pulses with small voltages, the crystalline Te device is robust in the OFF state, as no electric current can pass through the selector to program the memory unit below. When the voltage reaches a critical value, so that the accumulated Joule heating is sufficiently large to melt the crystalline state, the selector is switched to the ON state. After memory programming, and after the voltage is switched off, liquid Te is crystallized spontaneously, and the selector returns to the OFF state (*21*). Despite the simple monatomic composition, which avoids compositional fluctuation and segregation, Te-based selector devices still showed considerable noise in switching performance. A more in-depth understanding of the phase-transition processes in such devices is therefore required. However, in-situ atomic-scale experimental observation of nanosecond-level phase transitions is exceedingly difficult. Moreover, the atomic-level determination of crystallographic chirality makes such analyses even more challenging (*1*, *25*).

Herein, we reveal the atomic-level mechanisms of crystallization and chirality formation in α-Te. We performed quantum-accurate, ultra-large-scale atomistic simulations over millions of timesteps, directly comparable to the length (10–20 nm) and time scales (10–20 ns) of switching operations in Te-based selectors (*21*). To do so, we introduce a bespoke machine-learning (ML) interatomic potential model that reaches the accuracy of expensive meta-generalized gradient approximation (meta-GGA) density-functional-theory (DFT) computations, but at very much lower computational cost. We observed short-lived cubic-like motifs, which we show are key transient configurations in the crystallization of Te and in the formation of left- or right-handed chiral chains. Further annealing at higher temperatures showed grain-boundary ordering and chirality transfer, also mediated by disordered cubic-like motifs. Based on our atomic-scale simulations, we are able to explain the electrical-switching mechanism and behavior of Te-based selectors.

**The role of transient cubic-like motifs**

The origin of elemental crystal structures, including that of Te, has been discussed in the context of Peierls distortions: starting from a simple cubic structure, distortions and bond-breaking formally lead to different structure types (*26, 27*). Such three-dimensional Peierls distortions give *n* short bonds and *m* long bonds (with $n + m = 6$), yielding crystal structures of several pnictogen and chalcogen elements (Fig. S1). The chiral structure of α-Te can be viewed as a distorted cubic phase with a "2 + 4" bonding environment (Fig. 1): each atom takes part in two strong covalent bonds, forming the helical chains (*27*). We quantified the energy cost of forming a hypothetical cubic phase for various elements, relative to their ground-state crystal structures (Supplementary text and Fig. S1B), and found that a lower energy penalty for the simple cubic phase is correlated with faster crystallization for all of Sb (*28*), Bi (*29*), and Te (*18*). The easier formation of cubic-like motifs in Te was suggested to stem from delocalized lone-pair electrons and thus from an enhanced interchain bonding (*30*).

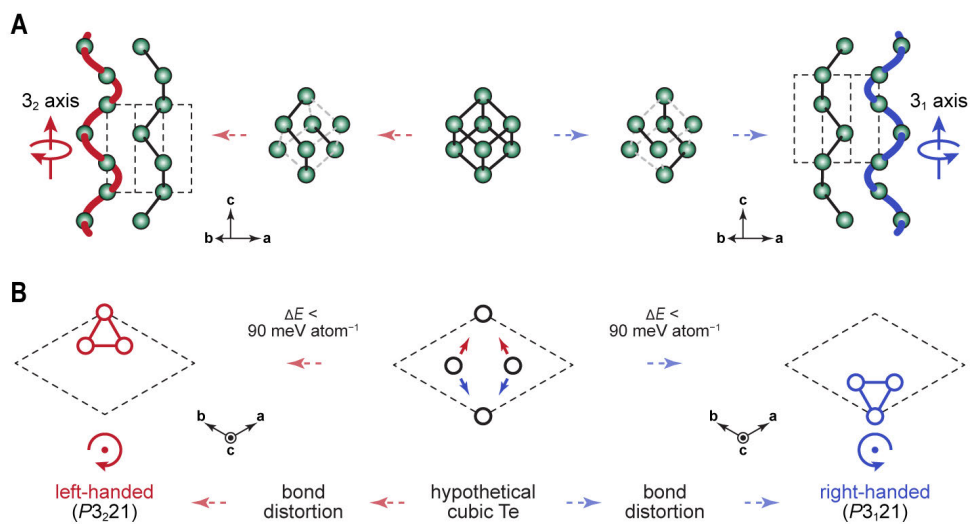

**Fig. 1. The microscopic pathway to chirality in Te crystals.** (**A**) The transitions from a simple cubic structure to the left-handed ($P3_221$) and right-handed ($P3_121$) crystal structure of Te, via bond breaking and distortion, sketched along the [120] direction. Dashed lines represent the unit cell of α-Te. The chiral helices formed by the left-handed ($3_2$) and right-handed ($3_1$) screw axes are emphasized by thick red and blue lines, respectively, drawn in a style similar to Ref. (*1*). (**B**) Top view of the crystallographic unit cells along the *c*-axis. The computed energy barrier of such transitions is less than 90 meV atom$^{-1}$ (details are given in the Supplementary text and Fig. S2).

Simply rotating or displacing the right-handed ($P3_121$) α-Te crystal without breaking chemical bonds cannot give the left-handed ($P3_221$) form, and vice versa. However, such chirality transfer is theoretically feasible by using the cubic phase as an intermediate, rearranging into a new 2 + 4 bonded motif that gives rise to a different chiral helix (Fig. 1). Such a transfer only requires very few atomic displacements, and the computed low-energy barrier supports the idea that chirality transfer via cubic-like motifs might happen at finite temperatures (Fig. S2). We found no saddle points between the ground state and a hypothetical cubic Te phase, suggesting that disordered cubic-like Te is energetically unstable (i.e., not a metastable phase).

**Crystallization of Te at the atomic scale**

We ran large-scale ML-driven molecular-dynamics (MLMD) simulations—for 102,000 atoms in a 15.2 × 15.5 × 15.0 nm$^3$ box over tens of nanoseconds—to study the crystallization of supercooled amorphous Te. The development and validation of the ML interatomic potential model for Te, using a domain-specific iterative training protocol (*31*) and the atomic cluster expansion (ACE) framework (*32*) with optimized hyperparameters (*33*), are described in the supplementary material. Notably, we included the transition path between cubic and chiral Te in the training database, to capture the subtle structural changes occurring during crystallization. Figure 2A shows one such crystallization simulation; in all, 5 independent runs were performed (Fig. S3). We used the Smooth Overlap of Atomic Positions (SOAP) similarity measure to quantify the degree of "per-atom crystallinity", $\bar{k}$ (*34*, *35*), with respect to ideal α-Te, allowing us to distinguish atoms in amorphous-like ($\bar{k} < 0.75$) and crystal-like ($\bar{k} \geq 0.75$) regions in the simulated model. We further separated the crystal-like atoms into two types: disordered cubic-like motifs ($0.75 \leq \bar{k} < 0.85$) with bond angles close to 90°; and chain-like atoms ($\bar{k} \geq 0.85$) that clearly belong to either left- or right-handed helices. Both disordered cubic-like and chain-like atoms were found in supercooled amorphous Te (Fig. S4) (*36*, *37*), indicating similar local bonding environments in the amorphous and crystalline forms.

The start of the crystallization process (≈1 ns into the simulation) was signaled by the creation of a crystal-like seed containing tens of the over-coordinated cubic-like atoms and no chain-like atoms (Fig. 2A). After that, more cubic-like seeds appeared, indicating a nucleation process with multiple nuclei, consistent with experimental observations at room temperature (*18*). The potential energy dropped markedly (Fig. 2B) and the numbers of atoms in cubic-like motifs increased (Fig. 2C) during incubation (0–1 ns) and nucleation (1–3 ns) periods. At ≈3 ns, the cubic-like precursors suddenly underwent swift and synergistic distortions, forming chain-like fragments with only two primary bonds (2 + 4) and near-103° bond angles (Fig. 2A and S5). Such fast and collective behavior was also evidenced by the abrupt increase of two-coordinated atoms at ≈3 ns (Fig. S6). The estimated lifetime of the cubic-like precursors is ≈1 ns, based on the different onsets of forming disordered cubic-like and chain-like motifs (Fig. 2C inset).

Following nucleation, fast growth was observed from 3–5 ns, concomitant with a linear decrease in potential energy (Fig. 2B) and the increasing presence of crystal-like environments (Fig. 2C). Notably, disordered cubic-like motifs of atoms still appeared at the crystalline–amorphous interfaces at the growth fronts, whereas atoms inside the grains assembled into chiral chains (Fig. 2A and Fig. S7). The cubic-like motifs facilitate inter-chain bonding, leading to nearly isotropic growth (rather than anisotropic growth along the chain direction only). In contrast to the sudden and collective transition from cubic- to chain-like fragments in nucleation, the growth proceeded atom-by-atom, similar to the growth mechanism of Ge–Sb–Te phase-change materials (*34*, *38*). The resultant polycrystalline model has ≈80% of crystal-like atoms (Fig. 2C). We note that the recrystallized structure in our simulations is consistent with atomic-level images observed in experiments (*21*) and in recent small-scale *ab initio* molecular-dynamics simulations of templated crystal growth (*30*), but is very different from the icosahedral and body-centered-cubic like clusters found in previous simulations based on an empirical interatomic force field (*39*).

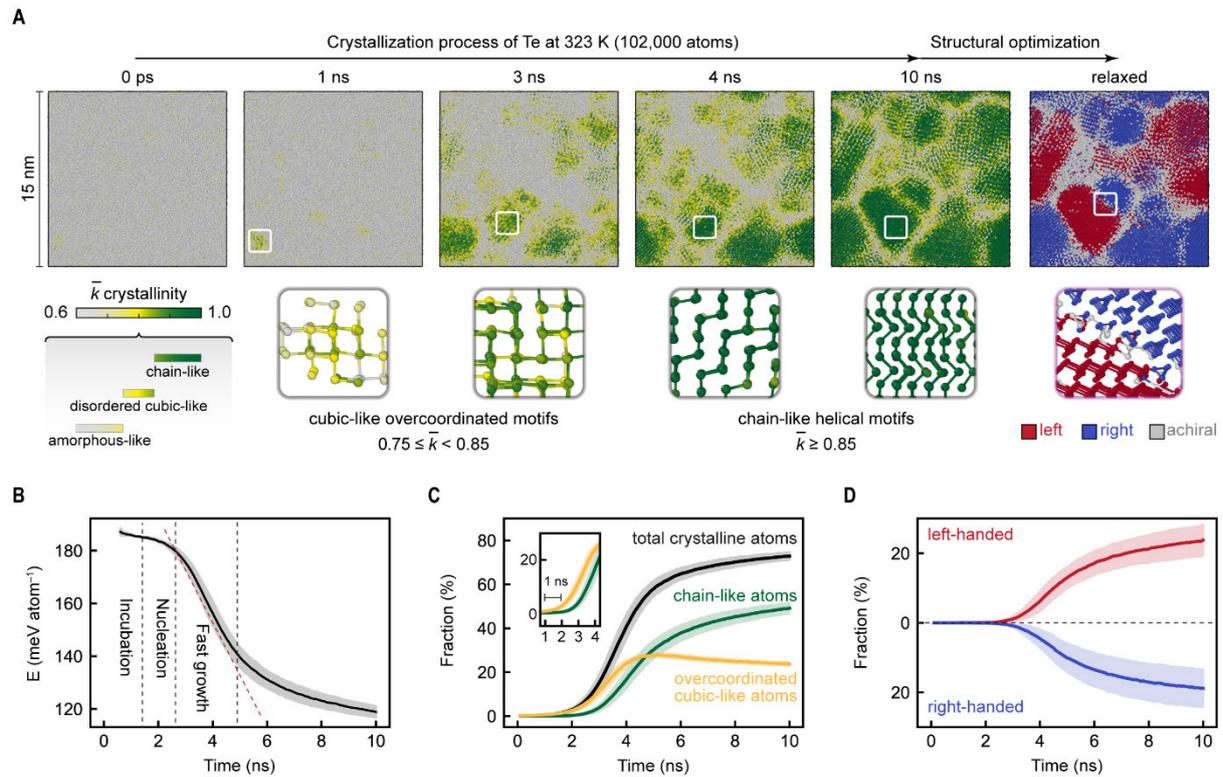

**Fig. 2. Nucleation and crystallization of Te.** (**A**) Crystallization of supercooled Te from ML-driven molecular dynamics, at 323 K for 10 ns. The structural model contains 102,000 atoms. Atoms are color-coded using the per-atom SOAP crystallinity $\bar{k}$, illustrating the gradual structural ordering from the (disordered) supercooled liquid phase to the polycrystal. Atoms in the rightmost panel are color-coded according to chirality. The lower panels highlight typical local structural motifs during the crystallization process. (**B**) Energy profile of the crystallization process, divided into three different stages, viz. incubation, nucleation, and fast growth. The energy is given relative to that of trigonal Te. (**C**) Evolution of the fraction of chain-like and cubic-like motifs. The inset shows different onsets for forming cubic-like and chain-like motifs. (**D**) The counts of atoms in left- and right-handed helices formed during the crystallization. All data in (**B**)–(**D**) are averaged over 5 independent samples (500 snapshots in total), and the shaded areas indicate the standard deviation at each timestep for the 500 snapshots.

We defined a measure for the per-atom chirality, based on local geometry and helical connectivity, and classify atoms as left-handed, right-handed, or achiral (Supplementary text). We found a random distribution of left- (red) and right-handed (blue) chiral grains. Two chiral grains with different chain directions were connected by a disordered boundary (Fig. 2A). Averaged over five independent samples, the numbers of left- and right-handed chiral-environment atoms during crystallization are very similar (Fig. 2D), consistent with the fact that the routes from the hypothetical cubic phase to left- and right-handed helices are energetically equivalent (Fig. 1).

**Grain-boundary ordering and chirality transfer**

Looking beyond the formation of chiral chains, excess kinetic energy at higher temperatures likely results in direct chirality transfer between left- and right-handed chains (Fig. 1B). To reveal the mechanism of such chirality transfer, we annealed the crystallized structural model at higher temperatures, from 373 to 473 K (Fig. S8). Figure 3A shows marked grain-boundary ordering and chirality transfer in polycrystalline Te at 473 K. Early on during annealing, small chiral grains were consumed by neighboring larger grains. According to classical nucleation theory, the annihilation of small unstable grains stems from their large interfacial energies compared to the bulk. The surviving chiral grains continuously compete with each other, leading to frequent chirality transfer and changes in grain boundaries. We note that these effects are driven by the formation of energetically more favorable interfaces between neighboring grains, as evidenced by the ML per-atom energies (Fig. S9). After annealing, we found only two chiral grains, one left- and one right-handed (Fig. 3A).

Notably, these two grains had the same chain direction, nearly perpendicular to the $z$-axis of the simulation box, and were separated by sharp boundaries. We color-coded the well-annealed structural model by per-atom chirality and $\bar{k}$ crystallinity similarity (Fig. 3B), showing that the chiral grain boundaries are highly crystal-like and energetically stable (Fig.

S10). Both chiral grains show a pyramid-like morphology with sharp facets (Fig. S11). These observations from our atomistic simulations are consistent with the 3D chiral morphology characterized experimentally using STEM (*9*). Moreover, we observed screw dislocations at the boundary of one chiral grain (Fig. S10), in agreement with atomic-resolution STEM (*9*), which have been proposed to mediate the formation of morphographic chirality by creating a more reactive crystal-growth front (*40–43*). Taken together, this evidence implies an atomic-level origin of morphographic chirality—likely due to forming screw dislocations and stable facets during growth and ordering—instead of resulting from crystallographic chirality alone.

We next looked into atomistic routes of chirality transfer, and found two different types (Fig. 3C–D). The first type occurred at the boundary between two chiral grains with different chain directions, proceeding via gradual atomic rearrangements at the interfaces. Figure 3C shows a representative transition process: many amorphous-like atoms ($\bar{k} < 0.75$) were found at a disordered boundary between two adjacent grains. Given excess kinetic energy, these atoms rearranged and formed energetically more favorable cubic-like motifs. We found a marked similarity between this chirality transfer and the growth process (Fig. S7). Both quickly proceeded at the interface, via formation of disordered cubic-like motifs, leading to chiral fragments. The grain-boundary ordering was mostly dominated by this growth-type chirality transfer, rapidly reducing the number of grains (Fig. S12).

The second type of chirality transfer proceeded in a more collective way. Due to considerable thermal fluctuations, chiral atoms inside crystal grains might temporarily become achiral but remain in cubic-like motifs (cf. Fig. 1). In analogy to the collective structural transition to chiral helices in nucleation, such cubic-like fragments were not stable, and they either returned to the original chiral structures or formed other chiral helices with different handedness or chain directions (Fig. 3D). We found a remarkable appearance of this nucleation-type chirality transfer at 13–14 ns, after two large grains connected via sharp boundaries, resulting in a sudden change in the left-to-right-handed ratio (Fig. S12). In fact, we also observed a change in chain direction during small-scale growth simulation with a prefixed crystalline template using ab initio molecular dynamics (Fig. S7), also mediated by the formation of cubic-like motifs. In our MLMD simulations, all disordered grain boundaries were eliminated at the end of the reordering, and no chirality transfer was evident in the stable chiral grains.

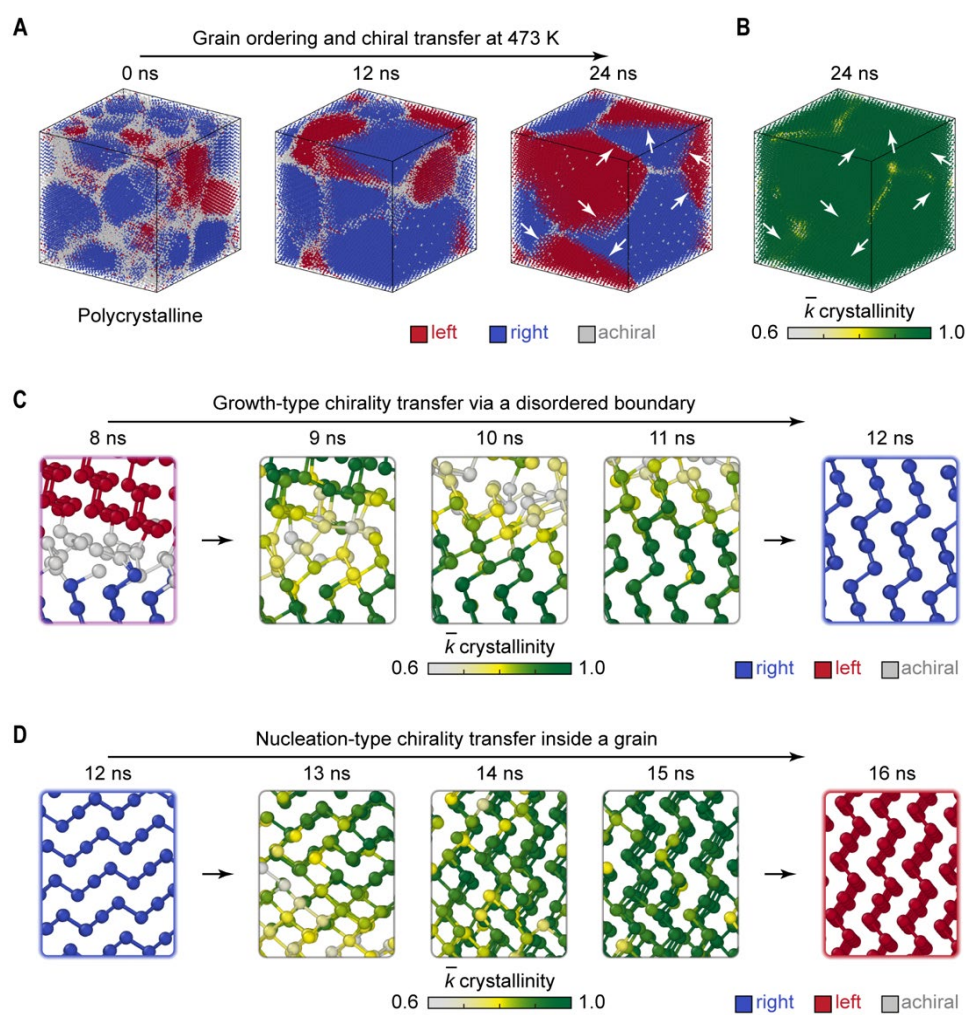

**Fig. 3. Grain-boundary ordering and chirality transfer in Te.** (**A**) Annealing the crystallized Te model (Fig. 2A) at 473 K for another 24 ns. Atoms are color-coded according to their per-atom chirality, indicating frequent chirality transfers during annealing. The resultant structure contains only two chiral grains under periodic boundary conditions; white arrows indicate the sharp boundaries between the grains. (**B**) The fully annealed structure, color-coded using the $\bar{k}$ crystallinity measure, in the same way as shown in Fig. 2A. White arrows indicate the boundary between two chiral grains. (**C**, **D**) Typical snapshots showcasing two different types of chirality transfer during the annealing process: (**C**) growth-type and (**D**)

nucleation-type chirality transfer. Atoms are color-coded using per-atom SOAP crystallinity $\bar{k}$ or per-atom chirality.

**Switching in Te selector devices**

Finally, we clarify why there is a crucial difference in the switch-on voltage between the first switching cycle and all subsequent switching cycles, and reveal the origin of the large noise observed in Te selector devices (*21*). Figure 4A shows typical schematic experimental current–voltage data of Te selector devices in the first switching cycle, sketched after Refs. (*21*, *44*), and this large switch-on voltage is denoted as $V_{\text{fire}}$ (2.5–3.5 V). The higher the initial annealing temperature, the larger the $V_{\text{fire}}$ value. In the subsequent switches, only a small threshold voltage, denoted as $V_{\text{th}}$, was required to set the ON state (1.0–1.5 V). We performed full-loop phase-transition simulations using a structural model of approximately 15 × 15 × 15 nm$^3$ (containing 102,000 Te atoms), comparable to real devices with 10–20-nm thick Te films (*21*). We first carried out a 5-ns heating simulation on an α-Te supercell model, representing the idealized case of starting from a perfect single crystal (Fig. 4B). The structure did not melt until reaching the experimental melting temperature of 723 K (Fig. 4C). We then took the melted structure from the first cycle, and cooled it from 800 to 300 K within 5 ns. We kept the structural model at 300 K for another 12 ns until it was fully crystallized (Fig. S13). The whole crystallization process was consistent with the observations in Fig. 2A, resulting in a polycrystalline state of Te with lots of chiral grains after the first switch (Fig. 4D). Next, we heated this model for another 5 ns to enter the ON state again (Fig. 4E). In contrast to the melting process of an α-Te single crystal (Fig. 4B), the polycrystalline structure started to disorder via the grain boundaries at 523 K (Fig. 4F)—much lower than the normal melting temperature of Te (723 K). Hence, given the presence of abundant grain boundaries as melting "seeds" in the polycrystal after nanosecond crystallization, the $V_{\text{th}}$ required for the second switch is expected to be much smaller than the $V_{\text{fire}}$ needed to melt a single crystal.

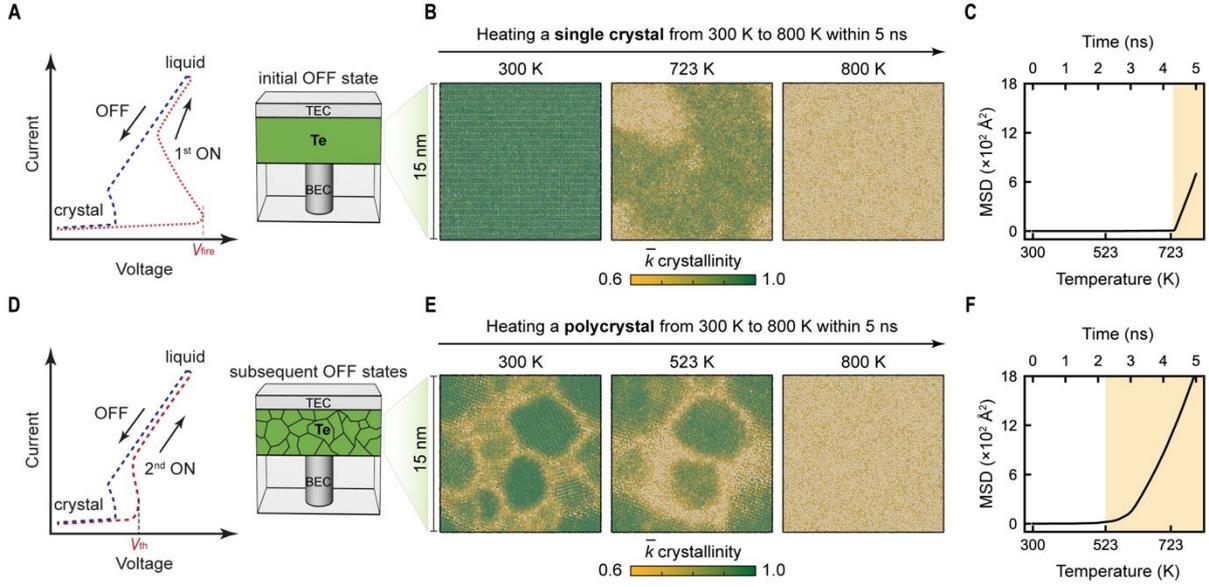

**Fig. 4. Device-scale switching process of Te.** (**A**) A schematic experimental current–voltage curve during the first switch of Te selectors (*21*), in which the initial state is a single crystal. Inset: sketch of Te-based selectors in memory devices reported in Ref. (*21*). The Te film, with a thickness of 10 to 20 nm, is sandwiched by top and bottom electrode contacts (TEC and BEC). (**B**) A 5-ns simulated heating process (from 300 to 800 K) of single-crystalline Te. Atoms are color-coded by per-atom SOAP crystallinity $\bar{k}$, similar to Fig. 2A. (**C**) The calculated mean-square displacement (MSD) of atoms during the heating process. (**D**) A schematic experimental current–voltage curve during the second switch of Te selectors (*21*), in which the Te film is polycrystalline. (**E, F**) As panels (B, C), but now for a polycrystalline sample, showing distinctly faster melting.

We also examined five crystallization simulations with independent thermal histories, and found chiral grains of different size, orientation, and handedness in the recrystallized structures (Fig. S3). We post-annealed one of the recrystallized structures at higher temperatures (from 323 to 473 K) for another 24 ns (Fig. S8), and took the deeply annealed structures for subsequent melting simulations (Fig. S14). Clearly, a higher melting temperature is needed for the polycrystalline structure with fewer grain boundaries: the structural disordering only occurred at the sharp grain boundaries at 653 K in the crystallized model with two chiral grains, and then expanded across the entire structural model (Fig. S15). Hence, the increased switch-on voltage upon thermal annealing is a direct consequence of grain numbers and sizes in the OFF-state structure, consistent with the qualitative arguments from Ref. (*45*). For the cross-point memory integration, it is important to have a relatively large switch-on voltage (> 3 V) for the selector units to guarantee the programming window for the memory units (*46*). Therefore, additional post-annealing pulses could be applied to reduce the structural stochasticity of the OFF-state before the subsequent switch-on operations, which could simultaneously improve the programming consistency and increase the $V_{th}$ value for practical use.

**Conclusions and outlook**

We have revealed the microscopic mechanisms of crystallization and chirality formation in elemental Te, one of the simplest chiral inorganic systems, based on quantum-accurate and realistic-scale ML-driven simulations. Instead of directly forming helical chains, the rapid nucleation in Te starts with higher-coordinated, disordered cubic-like seeds. With only nanosecond lifetimes, these transient motifs undergo fast and collective distortions to form chiral helices. We found two different types of chirality transfer during high-temperature annealing, resulting in a marked grain-boundary-ordering process, also mediated by disordered cubic-like motifs. Our simulations allowed us to explain the switching mechanisms of Te-based selector devices on a fundamental level. We found that the degree of polycrystallinity is key to determining the switching voltage—the fewer grains there are in the SET state, the larger this voltage. Our work has revealed the atomic-level origins of chirality in α-Te and can provide a starting point for realistic computational studies of other chiral crystals.

**Acknowledgments**

We thank Dr. Yuanbin Liu for help with fitting the ML potentials. Y.Z. acknowledges a China Scholarship Council-University of Oxford scholarship. S.R.E. acknowledges the Leverhulme Trust (UK) for a Fellowship. W.Z. thanks the support by the National Key Research and Development Program of China (2023YFB4404500), the National Natural Science Foundation of China (62374131) and the International Joint Laboratory for Micro/Nano Manufacturing and Measurement Technologies of XJTU. V.L.D. acknowledges a UK Research and Innovation Frontier Research grant [grant number EP/X016188/1]. This work was performed using resources provided by the Cambridge Service for Data Driven Discovery (CSD3) operated by the University of Cambridge Research Computing Service (www.csd3.cam.ac.uk), provided by Dell EMC and Intel using Tier-2 funding from the Engineering and Physical Sciences Research Council (capital grant EP/T022159/1), and DiRAC funding from the Science and Technology Facilities Council (www.dirac.ac.uk). We are grateful for computational support from the UK national high-performance computing service, ARCHER2, for which access was obtained via the UKCP consortium and funded by EPSRC grant ref EP/X035891/1, as well as through a separate EPSRC Access to High-Performance Computing award.

**Data and materials availability**

Data supporting this work, including the parameter files required to use the potential, the fitting data, and the structural models, will be made openly available in a suitable repository upon journal publication.

**References**

1. Z. Dong, Y. Ma, Atomic-level handedness determination of chiral crystals using aberration-corrected scanning transmission electron microscopy. *Nat. Commun.* **11**, 1588 (2020).

2. A. Ben-Moshe, S. G. Wolf, M. B. Sadan, L. Houben, Z. Fan, A. O. Govorov, G. Markovich, Enantioselective control of lattice and shape chirality in inorganic nanostructures using chiral biomolecules. *Nat. Commun.* **5**, 4302 (2014).

3. A. G. Shtukenberg, Y. O. Punin, A. Gujral, B. Kahr, Growth actuated bending and twisting of single crystals. *Angew. Chem. Int. Ed.* **53**, 672–699 (2014).

4. M. C. di Gregorio, L. J. W. Shimon, V. Brumfeld, L. Houben, M. Lahav, M. E. van der Boom, Emergence of chirality and structural complexity in single crystals at the molecular and morphological levels. *Nat. Commun.* **11**, 380 (2020).

5. P. Wang, S.-J. Yu, A. O. Govorov, M. Ouyang, Cooperative expression of atomic chirality in inorganic nanostructures. *Nat. Commun.* **8**, 14312 (2017).

6. D. Monego, S. Dutta, D. Grossman, M. Krapez, P. Bauer, A. Hubley, J. Margueritat, B. Mahler, A. Widmer-Cooper, B. Abécassis, Ligand-induced incompatible curvatures control ultrathin nanoplatelet polymorphism and chirality. *Proc. Natl. Acad. Sci. U. S. A.* **121**, e2316299121 (2024).

7. A. V. Gonzalez, M. Gonzalez, T. Hanrath, Emergence and inversion of chirality in hierarchical assemblies of CdS nanocrystal fibers. *Sci. Adv.* **9**, eadi5520 (2023).

8. S. W. Im, D. Zhang, J. H. Han, R. M. Kim, C. Choi, Y. M. Kim, K. T. Nam, Investigating chiral morphogenesis of gold using generative cellular automata. *Nat. Mater.* **23**, 977–983 (2024).

9. A. Ben-Moshe, A. da Silva, A. Müller, A. Abu-Odeh, P. Harrison, J. Waelder, F. Niroui, C. Ophus, A. M. Minor, M. Asta, W. Theis, P. Ercius, A. P. Alivisatos, The chain of chirality transfer in tellurium nanocrystals. *Science* **372**, 729–733 (2021).

10. K. Nakayama, A. Tokuyama, K. Yamauchi, A. Moriya, T. Kato, K. Sugawara, S. Souma, M. Kitamura, K. Horiba, H. Kumigashira, T. Oguchi, T. Takahashi, K. Segawa, T. Sato, Observation of edge states derived from topological helix chains. *Nature* **631**, 54–59 (2024).

11. G. Gatti, D. Gosálbez-Martínez, S. S. Tsirkin, M. Fanciulli, M. Puppin, S. Polishchuk, S. Moser, L. Testa, E. Martino, S. Roth, Ph. Bugnon, L. Moreschini, A. Bostwick, C. Jozwiak, E. Rotenberg, G. Di Santo, L. Petaccia, I. Vobornik, J. Fujii, J. Wong, D. Jariwala, H. A. Atwater, H. M. Rønnow, M. Chergui, O. V. Yazyev, M. Grioni, A. Crepaldi, Radial spin texture of the Weyl fermions in chiral tellurium. *Phys. Rev. Lett.* **125**, 216402 (2020).

12. F. Calavalle, M. Suárez-Rodríguez, B. Martín-García, A. Johansson, D. C. Vaz, H. Yang, I. V. Maznichenko, S. Ostanin, A. Mateo-Alonso, A. Chuvilin, I. Mertig, M. Gobbi, F. Casanova, L. E. Hueso, Gate-tuneable and chirality-dependent charge-to-spin conversion in tellurium nanowires. *Nat. Mater.* **21**, 526–532 (2022).

13. G. Jnawali, Y. Xiang, S. M. Linser, I. A. Shojaei, R. Wang, G. Qiu, C. Lian, B. M. Wong, W. Wu, P. D. Ye, Y. Leng, H. E. Jackson, L. M. Smith, Ultrafast photoinduced band splitting and carrier dynamics in chiral tellurium nanosheets. *Nat. Commun.* **11**, 3991 (2020).

14. L. Li, S. Zhao, W. Ran, Z. Li, Y. Yan, B. Zhong, Z. Lou, L. Wang, G. Shen, Dual sensing signal decoupling based on tellurium anisotropy for VR interaction and neuro-reflex system application. *Nat. Commun.* **13**, 5975 (2022).

15. Y. Yang, M. Xu, S. Jia, B. Wang, L. Xu, X. Wang, H. Liu, Y. Liu, Y. Guo, L. Wang, S. Duan, K. Liu, M. Zhu, J. Pei, W. Duan, D. Liu, H. Li, A new opportunity for the emerging tellurium semiconductor: making resistive switching devices. *Nat. Commun.* **12**, 6081 (2021).

16. Z. He, Y. Yang, J.-W. Liu, S.-H. Yu, Emerging tellurium nanostructures: controllable synthesis and their applications. *Chem. Soc. Rev.* **46**, 2732–2753 (2017).

17. S. Li, H. Zhang, H. Ruan, Z. Cheng, Y. Yao, F. Zhuge, T. Zhai, Programmable nucleation and growth of ultrathin tellurium nanowires via a pulsed physical vapor deposition design. *Adv. Funct. Mater.* **33**, 2211527 (2023).

18. C. Zhao, H. Batiz, B. Yasar, H. Kim, W. Ji, M. C. Scott, D. C. Chrzan, A. Javey, Tellurium single-crystal arrays by low-temperature evaporation and crystallization. *Adv. Mater.* **33**, e2100860 (2021).

19. H. Keller, J. Stuke, Elektrische und optische Eigenschaften von amorphem Tellur. *Phys. Status Solidi B* **8**, 831–840 (1965).

20. A. S. Epstein, H. Fritzsche, K. Lark-Horovitz, Electrical properties of tellurium at the melting point and in the liquid state. *Phys. Rev.* **107**, 412–419 (1957).

21. J. Shen, S. Jia, N. Shi, Q. Ge, T. Gotoh, S. Lv, Q. Liu, R. Dronskowski, S. R. Elliott, Z. Song, M. Zhu, Elemental electrical switch enabling phase segregation-free operation. *Science* **374**, 1390–1394 (2021).

22. N. Gong, W. Chien, Y. Chou, C. Yeh, N. Li, H. Cheng, C. Cheng, I. Kuo, C. Yang, R. Bruce, A. Ray, L. Gignac, Y. Lin, C. Miller, T. Perri, W. Kim, L. Buzi, H. Utomo, F. Carta, E. Lai, H. Ho, H. Lung, M. BrightSky, "A no-verification multi-level-cell (MLC) operation in cross-point OTS-PCM" in *2020 IEEE Symposium on VLSI Technology* (2020), pp. 1–2.

23. A. Fazio, "Advanced technology and systems of cross point memory" in *2020 IEEE International Electron Devices Meeting (IEDM)* (2020), p. 24.1.1-24.1.4.

24. D. Kau, S. Tang, I. V. Karpov, R. Dodge, B. Klehn, J. A. Kalb, J. Strand, A. Diaz, N. Leung, J. Wu, S. Lee, T. Langtry, K. Chang, C. Papagianni, J. Lee, J. Hirst, S. Erra, E. Flores, N. Righos, H. Castro, G. Spadini, "A stackable cross point Phase Change Memory" in *2009 IEEE International Electron Devices Meeting (IEDM)* (2009), pp. 1–4.

25. Y. Ma, P. Oleynikov, O. Terasaki, Electron crystallography for determining the handedness of a chiral zeolite nanocrystal. *Nat. Mater.* **16**, 755–759 (2017).

26. J. K. Burdett, T. J. McLarnan, A study of the arsenic, black phosphorus, and other structures derived from rock salt by bond-breaking processes. I. Structural enumeration. *J. Chem. Phys.* **75**, 5764–5773 (1981).

27. A. Decker, G. A. Landrum, R. Dronskowski, Structural and electronic Peierls distortions in the elements (A): The crystal structure of tellurium. *Z. Anorg. Allg. Chem.* **628**, 295–302 (2002).

28. M. Salinga, B. Kersting, I. Ronneberger, V. P. Jonnalagadda, X. T. Vu, M. Le Gallo, I. Giannopoulos, O. Cojocaru-Mirédin, R. Mazzarello, A. Sebastian, Monatomic phase change memory. *Nat. Mater.* **17**, 681–685 (2018).

29. O. Hunderi, Optical properties of crystalline and amorphous bismuth films. *J. Phys. F: Met. Phys.* **5**, 2214 (1975).

30. Y. Sun, B. Li, T. Yang, Q. Yang, H. Yu, T. Gotoh, C. Shi, J. Shen, P. Zhou, S. R. Elliott, H. Li, Z. Song, M. Zhu, Nanosecond phase-transition dynamics in elemental tellurium. *Adv. Funct. Mater.*, DOI: 10.1002/adfm.202408725 (2024).

31. Y. Zhou, W. Zhang, E. Ma, V. L. Deringer, Device-scale atomistic modelling of phase-change memory materials. *Nat. Electron.* **6**, 746–754 (2023).

32. R. Drautz, Atomic cluster expansion for accurate and transferable interatomic potentials. *Phys. Rev. B* **99**, 014104 (2019).

33. D. F. Thomas du Toit, Y. Zhou, V. L. Deringer, Hyperparameter Optimization for Atomic Cluster Expansion Potentials. arXiv:2408.00656 [Preprint] (2024).

34. Y. Xu, Y. Zhou, X. D. Wang, W. Zhang, E. Ma, V. L. Deringer, R. Mazzarello, Unraveling crystallization mechanisms and electronic structure of phase-change materials by large-scale ab initio simulations. *Adv. Mater.* **34**, 2109139 (2022).

35. A. P. Bartók, R. Kondor, G. Csányi, On representing chemical environments. *Phys. Rev. B* **87**, 184115 (2013).

36. T. Ichikawa, Electron diffraction study of the local atomic arrangement in amorphous tellurium films. *Phys. Status Solidi B* **56**, 707–715 (1973).

37. J. Akola, R. O. Jones, Structure and dynamics in amorphous tellurium and Te$_n$ clusters: A density functional study. *Phys. Rev. B* **85**, 134103 (2012).

38. I. Ronneberger, W. Zhang, H. Eshet, R. Mazzarello, Crystallization properties of the Ge$_2$Sb$_2$Te$_5$ phase-change compound from advanced simulations. *Adv. Funct. Mater.* **25**, 6407–6413 (2015).

39. H. G. Abbas, J. R. Hahn, Crystallization mechanism of liquid tellurium from classical molecular dynamics simulation. *Mater. Chem. Phys.* **240**, 122235 (2020).

40. J. Zhu, H. Peng, A. F. Marshall, D. M. Barnett, W. D. Nix, Y. Cui, Formation of chiral branched nanowires by the Eshelby Twist. *Nat. Nanotechnol.* **3**, 477–481 (2008).

41. Y. Liu, J. Wang, S. Kim, H. Sun, F. Yang, Z. Fang, N. Tamura, R. Zhang, X. Song, J. Wen, B. Z. Xu, M. Wang, S. Lin, Q. Yu, K. B. Tom, Y. Deng, J. Turner, E. Chan, D. Jin, R. O. Ritchie, A. M. Minor, D. C. Chrzan, M. C. Scott, J. Yao, Helical van der Waals crystals with discretized Eshelby twist. *Nature* **570**, 358–362 (2019).

42. S. A. Morin, M. J. Bierman, J. Tong, S. Jin, Mechanism and kinetics of spontaneous nanotube growth driven by screw dislocations. *Science* **328**, 476–480 (2010).

43. M. J. Bierman, Y. K. A. Lau, A. V. Kvit, A. L. Schmitt, S. Jin, Dislocation-driven nanowire growth and Eshelby twist. *Science* **320**, 1060–1063 (2008).

44. M. Zhu, K. Ren, Z. Song, Ovonic threshold switching selectors for three-dimensional stackable phase-change memory. *MRS Bull.* **44**, 715–720 (2019).

45. X.-D. Wang, W. Zhang, E. Ma, Monatomic phase-change switch. *Sci. Bull.* **67**, 888–890 (2022).

46. H.-Y. Cheng, F. Carta, W.-C. Chien, H.-L. Lung, M. J. BrightSky, 3D cross-point phase-change memory for storage-class memory. *J. Phys. D: Appl. Phys.* **52**, 473002 (2019).

Supplementary Materials for

# The pathway to chirality in elemental tellurium

Yuxing Zhou[1], Stephen R. Elliott[2], Daniel F. Thomas du Toit[1], Wei Zhang[3*],
Volker L. Deringer[1*]

*Corresponding authors: wzhang0@mail.xjtu.edu.cn; volker.deringer@chem.ox.ac.uk

**Computational Methods**

<u>**Machine-learning-based interatomic potentials**</u>

We initially fitted an ML interatomic potential for tellurium based on the Gaussian approximation potential (GAP) framework (*S1*). The training dataset was generated using a domain-knowledge-based iterative protocol. We have shown in previous work (*S2*) that such a protocol produces reliable ML potentials, enabling the accurate description of chemically and structurally complex Ge–Sb–Te alloys under realistic device conditions. Highly converged DFT calculations were performed to obtain the energies and forces of the training structures (i.e., for the labeling of the reference dataset). To further accelerate the MD simulations, we employed the atomic cluster expansion (ACE) framework (*S3*), which is faster than GAP by at least two orders of magnitude (*S4*).

We fitted an ACE model based on the reference dataset of the GAP model using XPOT's interface (*S5*, *S6*) to PACEmaker (*S3*, *S4*, *S7*), facilitating hyperparameter optimization. For this work, we used hyperparameters optimized for a related material from an ongoing project that has not yet been published. The optimization resulted in a non-linear model with 3000 functions and three embeddings of the form:

$$E_i = \varphi + 4.6\sqrt{\varphi} + 0.1\varphi^2$$

where $E_i$ is the energy of atom $i$ and $\varphi$ is defined as the atomic property of atom $i$.

The ACE model was fitted without any structural weighting, i.e., each structure has the same weighting in the loss function minimization during fitting. We used an upfitting approach for fitting the ACE potential. This consisted of two separate fits of the potential, where the initial fit focused on converging the forces using $\kappa = 0.8$, and the second fit (upfit) improved the energy predictions by using $\kappa = 0.02$, where $\kappa$ controls the relative weighting of the energy and force errors in the loss function. The protocol used here is an updated version of the upfitting protocol used in Ref. (*S6*). The initial fit used a power-order ladder-fitting protocol, as defined in Ref. (*S7*), with a ladder step of 1000 functions, for a total of 2250 (750 × 3) iterations. The upfit did not use ladder fitting (this is not possible when fitting atop an existing potential) and employed 2000 iterations.

Both models provided accurate results as compared to DFT references and experimental data (Supplementary text). Notably, our ML models can capture the energetic difference between cubic-like and chain-like motifs (Fig. S2), with the largest predicted error being less than 20 meV atom$^{-1}$ as compared to DFT, which is key to revealing the transient cubic-like motifs.

Unlike DFT, which predicts the total energy of a given structure, ML potentials provide per-atom energies (*S1*, *S8*), and the total energy of a structure is the sum of the energies of each atom. As such, we can probe these local energies to investigate local energetics in materials. GAP local energies have been used to provide insightful energetic information in previous work, such as the energy distributions of disordered graphene (*S9*), coordination defects in amorphous (a-) Si (*S10*), and the local stabilization of simple-hexagonal (sh) crystallites during the pressurization of a-Si (*S11*). Given this successful track record of using GAP local energies to study local structure, our GAP model was used to compute the total energy and per-atom energies of the structural models, with the help of the package QUIP and its python interface quippy.

**<u>Molecular-dynamics simulations</u>**

We carried out ML-driven MD (MLMD) simulations using LAMMPS (*S12*) with an interface to the pacemaker package, which is available freely for academic non-commercial research at https://github.com/ICAMS/python-ace. We used the canonical ensemble (i.e., constant volume and temperature; NVT) in all ML-driven MD simulations. In this regard, the density of the structural model was fixed at the theoretical crystal density (i.e., 6.10 g cm$^{-3}$), computed at the TPSS level, which is slightly smaller than the experimental crystal density (i.e., 6.23 g cm$^{-3}$) (*S13*). The fixed volume is relevant to the restrained geometrical conditions imposed by the surrounding materials in confined Te-based selector devices. The timestep was set to 1 fs for all ML-driven MD simulations.

**DFT computations**

To label the reference dataset, the energies and forces were computed from single-point calculations using VASP (*S14*, *S15*). Projector augmented-wave (PAW) pseudopotentials were used (*S15*, *S16*). The meta-generalized gradient approximation (meta-GGA) type Tao–Perdew–Staroverov–Scuseria (TPSS) functional (*S17*) was employed, which has been demonstrated to enable accurate AIMD simulations of liquid and amorphous Te (*S18*, *S19*). The cut-off of the plane-wave expansion was set to 650 eV. In the single-point calculations, we used an energy convergence criterion of $10^{-7}$ eV per atom. A *k*-point grid with a spacing of 0.2 Å$^{-1}$ was used to sample reciprocal space.

The AIMD simulations shown in the present work were performed using the "second-generation" Car–Parrinello scheme (*S20*), as implemented in the Quickstep package of CP2K (*S21*). The Kohn–Sham orbitals were expanded using a triple-zeta plus polarization Gaussian-type basis set and we used plane waves with a cut-off of 300 Ry to expand the charge density. We used scalar-relativistic Goedecker pseudopotentials (*S22*) and the TPSS functional (*S17*). The Brillouin zone was sampled only at Γ, and the timestep was 2 fs.

**Visualization**

Structural models were visualized using OVITO (*S23*).

**Supplementary Text**

The role of transient cubic-like motifs

As mentioned in the main text, bond breaking and distortions from a perfect simple-cubic structure result in different structure types (*S24*, *S25*), including A8 (Fig. 1A), A17, and A7 (Fig. S1A). Such three-dimensional Peierls distortions give *n* short bonds and *m* long bonds (an "*n* + *m*" bonding environment, with *n* + *m* = 6), and thus lead to the ground-state crystal structures of many group-V (3 + 3) and group-VI (2 + 4) elements (Fig. S1A). Unlike layered rhombohedral Sb and black P, trigonal Te adopts a highly anisotropic and chiral crystal structure, which can be regarded as a distorted cubic phase with a 2 + 4 bonding environment (Fig. S1A).

We defined an energy indicator, the "cubic-phase formation penalty", $E_p$, expressed as the energy difference from a ground-state crystal structure to a (hypothetical) simple-cubic structure (i.e., $E_p = E_{\text{cubic-phase}} - E_{\text{ground-state}}$), for various pnictogen/chalcogen elements and chalcogenides (Fig. S1B). Such a quantitative indicator measures the energetic difficulty of forming a cubic-like phase. A gap at $E_p \approx 100$ meV atom$^{-1}$ clearly separates the lighter, third- and fourth-period elements (i.e., P, As, Se) from the heavier, fifth- and sixth-period elements (i.e., Sb, Bi, Te, Po). Notably, these two groups of elements have a different crystallization tendency. For the period-3 / -4 elements with $E_p > 100$ meV atom$^{-1}$, their crystallization temperatures are greatly above room temperature (*S26*, *S27*, *S28*), and the simple cubic phase is only observed under high pressures, e.g., in P (*S29*), As (*S30*), and Se (*S31*). By contrast, the period-5 / -6 elements with $E_p < 100$ meV atom$^{-1}$, e.g., Sb, Bi, and Te, show very fast crystallization, and can only remain in an amorphous state under nanoscale confinement (*S32*, *S33*, *S34*). It was also suggested that Sb forms a disordered cubic phase during fast nucleation in *ab initio* molecular-dynamics (AIMD) simulations (*S35*, *S36*). Moreover, many Ge–Sb–Te (GST) phase-change memory materials, with $E_p$ values of less than 100 meV atom$^{-1}$ (*S37*, *S38*), first crystallize into a metastable cubic phase within nanoseconds and then further into a stable hexagonal phase on high-temperature annealing (*S39*). Hence, we inferred that the easier formation of a cubic-like phase, which was suggested to result from the delocalized lone-pair electrons and thus an enhanced interchain bonding (*S40*), facilitates the formation of crystalline seeds in nucleation and thus leads to faster crystallization.

To evaluate the energy cost of forming a hypothetical cubic Te phase, we performed a grid search for the energetic landscape of different atomic configurations (Fig. S2): we fixed the symmetry and lattice parameters of the DFT-relaxed hexagonal unit-cell of trigonal Te, and tuned the relative positions of three atoms in the unit-cell. The three fractional coordinates are (0, 0, 0), ($x$, $y$, 1/3), and ($y$, $x$, 1/3), with $x, y \in (0, 1)$ and $x \geq y$. Based on the DFT-calculated energy profile, we found an energy barrier of less than 90 meV atom$^{-1}$ for forming a cubic-like motif. We also found direct transfer routes between left- and right-handed Te—a path that bypasses the idealized cubic phase and connects two chiral configurations (Fig. S2). Along the route, no saddle points were observed, suggesting that disordered cubic-like Te is not a metastable phase but merely a transient form.

Reference data and iterative training

To generate a reliable reference dataset covering a wide range of structural diversity in Te, we used a domain-knowledge-based iterative training protocol. To construct an initial reference dataset, which we called "iter-0", we first added isolated Te atom and Te–Te dimer configurations (in a large box of $20 \times 20 \times 20$ nm$^3$) in iter-0. Next, we carried out 8 rounds of GAP-based random structure search (GAP-RSS) processes to sample a wide range of random structural space (*S41*, *S42*). We also included the unit cells and supercells of crystal structures (containing less than 100 atoms) from the Materials Project database (*S43*), as well as various copies of these crystal structures with disordered lattice parameters and atomic coordinates. In addition, we included the configurations of liquid, supercooled liquid, and amorphous Te, as well as intermediate configurations during the growth simulations in iter-0, which were all obtained from AIMD simulations. Starting from the iter-0 dataset, we performed iterative training for GAP potentials, which can be divided into: (i) standard iterations that contain abundant supercooled liquid Te configurations at different temperatures and with varying atomic densities; and (ii) domain-specific iterations which include intermediate transition states during melting and crystallization of Te. The domain-specific iterations were specifically designed in this protocol so that our ML model can well describe the phase-transition process of Te, which is the key to this work. In total, 30 standard iterations and 20 domain-specific iterations were carried out in our iterative training process.

To obtain a stable GAP potential, we performed a screening of the original dataset discussed above and ruled out "unreasonable" configurations, including: (i) bulk structures with too small or too large atomic densities (i.e., $\rho < 0.015$ or $\rho > 0.045$ Å$^{-3}$); (ii) structures with too unfavorable energies ($E > 20$ eV per structure above the free atoms); and structures with too large force components ($|F_{ij}| > 20$ eV Å$^{-1}$, where $i$ runs over all atoms in the structure and $j$ represents force components along the $x$, $y$, and $z$ directions). We note that most of the removed configurations are RSS structures. By contrast, to fit a stable ACE potential, we kept the "unreasonable" configurations that were removed from the GAP model, but removed all Te–Te dimer configurations. Retaining the unreasonable configurations improved the robustness of the ACE potential without significantly reducing accuracy on lower energy structures. However, removing the dimers increased the prediction accuracy on the bulk Te testing dataset. Unlike GAP, ACE potentials do not have an explicit, separate two-body term, reducing the effectiveness of adding dimer configurations. A detailed summary of our training datasets for GAP and ACE models is provided in Table S1.

Validation of ML potentials

Our ML potentials were carefully validated, following guidelines in Ref. (*S44*). We first performed a numerical validation by computing the prediction errors based on an external dataset that was not included in the training reference dataset. The small root-mean-square errors (RMSE) of both predicted energies and forces indicated a good numerical accuracy of both ML models (Table S2). Next, we ran more comprehensive validations based on physical properties of Te. We computed the energy profiles of trigonal Te using our GAP and ACE models, and compared to the DFT reference (Fig. S2). We note that the energy difference of subtle structural changes between cubic-like and chain-like configurations can be well captured by our ML models, which is key to being able to describe crystallization and chiral transfer in Te. We also carried out structural validation on liquid, supercooled liquid, and amorphous Te structures from AIMD and MLMD, by computing the radial distribution function (RDF), angular distribution function (ADF), and angular-limited three-body function. We found that the results from GAP- and ACE-driven MD simulations were in good agreement with the AIMD reference (Fig. S16). We also computed the structure factor of liquid Te at various temperatures, and observed good consistency between our MLMD results and the experimental references (*S18*, *S45*) (Fig. S17).

Generation of structural models

Structural models for disordered (liquid, supercooled liquid, and amorphous) Te at different temperatures were generated following melt-quench protocols. Specifically, the initial structural model (e.g., hard-sphere random structure) was first randomized at 3,000 K for 30 ps. It was then cooled down to 1,000 K, above the melting point of Te, within 30 ps. The liquid model was equilibrated at 1,000 K for another 30 ps. The structural models were then further quenched down and held at different temperatures for another 30 ps, of which trajectories during the last 20 ps were taken for structural analysis. The canonical ensemble (NVT) was used with experimentally measured densities at different temperatures (*S18*, *S19*). The large-scale structural model of supercooled amorphous Te (which contains 102,000 atoms at a theoretical crystal density of 6.10 g cm$^{-3}$), as discussed in the main text, was also created via the same melt-quench process.

To generate structural models for growth simulations using AIMD (Fig. S7), we started with a supercell of idealized trigonal Te (containing 720 atoms at a theoretical crystal density of 6.10 g cm$^{-3}$). We pre-fixed three atomic layers of crystalline atoms in two crystal planes, i.e., the (001) and (210) planes. The melt-quench process was then performed to eliminate the ordering in the remaining part of the models. The fixed atoms were then released (and unfixed) in the growth simulations.

Quantification of per-atom crystallinity and chirality

The Smooth Overlap of Atomic Positions (SOAP) kernel was utilized to measure the similarity between two atomic environments (*S46*). A detailed discussion of this approach in the context of crystallinity was given in Ref. (*S47*). To quantify the per-atom crystallinity, we here set the reference structure to be α-Te, in which the lattice parameters and atomic coordinates have been fully optimized using DFT. The hyperparameters used in the SOAP-based crystallinity are $r_c$ = 9.0 Å, $\sigma_{at}$ = 0.3 Å, $l_{max}$ = 16, and $n_{max}$ = 16. The calculation of the SOAP power-spectrum vectors was carried out using the QUIP/quippy and GAP code (https://github.com/libAtoms/QUIP). We note that both amorphous-like and liquid-like atoms have relatively low per-atom crystallinity values. Hence, to clearly separate the crystallization from the melting process of Te, we used two different color schemes to represent the per-atom crystallinity in this work, as shown in Fig. 2 (i.e., the crystallization process) and Fig. 4 (i.e., the melting process) in the main text.

To quantify the per-atom chirality in the recrystallized Te structures, we defined a structural indicator, the chiral index, to describe the well-defined chiral shape of any given atom. A similar idea of identifying chirality in water molecules was given in Ref. (*S48*), which was based on the geometrical connections of neighboring molecules. The determination of per-atom chirality can be separated into three steps (Fig. S18).

Firstly, the per-atom crystallinity was computed for each atom. We note that only the crystal-like atoms (i.e., with a high similarity to the reference α-Te) have a chance to be a "chiral atom", and therefore the disordered atoms are all achiral. Moreover, the coordination number (CN) of a well-defined chiral Te atom can only be two. Hence, crystal-like atoms with incorrect coordination numbers (i.e., CN ≠ 2) were regarded as achiral atoms. An atomic cutoff of 3.2 Å was used to determine the coordination numbers.

Secondly, we examined the connection of a given atom $i$ to its two bonded atoms $j$ and $k$. In idealized α-Te, the bonding angle $\theta_{jik}$ (where atom $i$ is the central atom) should be around 103°. Given thermal fluctuations of atomic environments at finite temperatures, we manually set a threshold and ruled out unreasonable values of $\theta_{jik}$: namely, only atoms with $\theta_{jik} \in [80°, 160°]$ were considered as chiral fragments, because motifs with too large or too small angles largely deviate from the chiral fragments of trigonal Te.

Lastly, we determined the possible helix in which a given atom would take part. Given the connectivity of three atoms that satisfies the criteria in the first and second steps (viz. atoms $i, j$, and $k$, in which atom $i$ is bonded to $j$ and $k$), the atomic chirality is still ambiguous (cf. Fig. S18). Hence, a fourth atom $l$, which is bonded with either atom $j$ or atom $k$, is needed to fully define the per-atom chirality. Notably, we also ruled out the circumstance that these four atoms are nearly in the same plane based on the computed dihedral angles of these four atoms. We computed the vectors $\vec{v}_{ij}$ (with a direction from atom $i$ to atom $j$) and $\vec{v}_{ik}$ (with a direction from atom $i$ to atom $k$), and therefore the normal vector of the $ijk$ plane can be obtained as the cross product

$$\vec{v}_{\text{norm}} = \vec{v}_{ij} \times \vec{v}_{ik}.$$

Next, we checked the connectivity of the fourth atom, $l$. On one hand, if atom $l$ is connected to atom $j$, we can obtain another vector $\vec{v}_{lk}$ (with a direction from atom $l$ to atom $k$) which is parallel to the chain direction of this helical fragment. We then compute the dot product of $\vec{v}_{\text{norm}}$ and $\vec{v}_{lk}$,

$$c = \vec{v}_{\text{norm}} \cdot \vec{v}_{lk}.$$

If $c > 0$, then the fragment is left-handed; and if $c < 0$, then the fragment is right-handed (Fig. S18). On the other hand, if atom $l$ is connected to atom $k$, we can also have a vector $\vec{v}_{lj}$ (with a direction from atom $l$ to atom $j$). We compute the dot product of $\vec{v}_{\text{norm}}$ and $\vec{v}_{lj}$,

$$c = \vec{v}_{\text{norm}} \cdot \vec{v}_{lj}.$$

If $c > 0$, then the fragment is right-handed, and if $c < 0$, then the fragment is left-handed (Fig. S18).

Notably, the thermal fluctuation of atoms at finite temperatures results in considerable structural noise (e.g., local bond distortions). Hence, to better define the chirality of different chiral grains, we performed a conjugate-gradient structural optimization on the structural model using our ACE potential to remove the effect of thermal fluctuations. We performed this correction in Fig. 2A and Fig. 3 of the main text, as well as in Fig. S10 and Fig. S11.

To compute the number of grains in MD simulations (cf. Fig. S12), a graph-clustering algorithm described in Ref. (*S49*) was employed to separate chiral atoms with different chain orientations (i.e., grain segmentation). A graph was constructed based on any given atomic configuration: atoms were taken as nodes and chemical bonds were taken as edges. The edge weights of the graph were defined as: $W = e^{-\theta^2/3}$ [as also used in OVITO (*S23*)], in which $\theta$ is the angle between the chain direction (i.e., $\vec{v}_{lk}$ or $\vec{v}_{lj}$ in Fig. S18) and the $c$-axis (i.e., the [001] direction). We performed the grain segmentation only on left- and right-handed grains, but separately. In other words, grains with different chiral shape but the same chain directions are regarded as two different (chiral) grains.

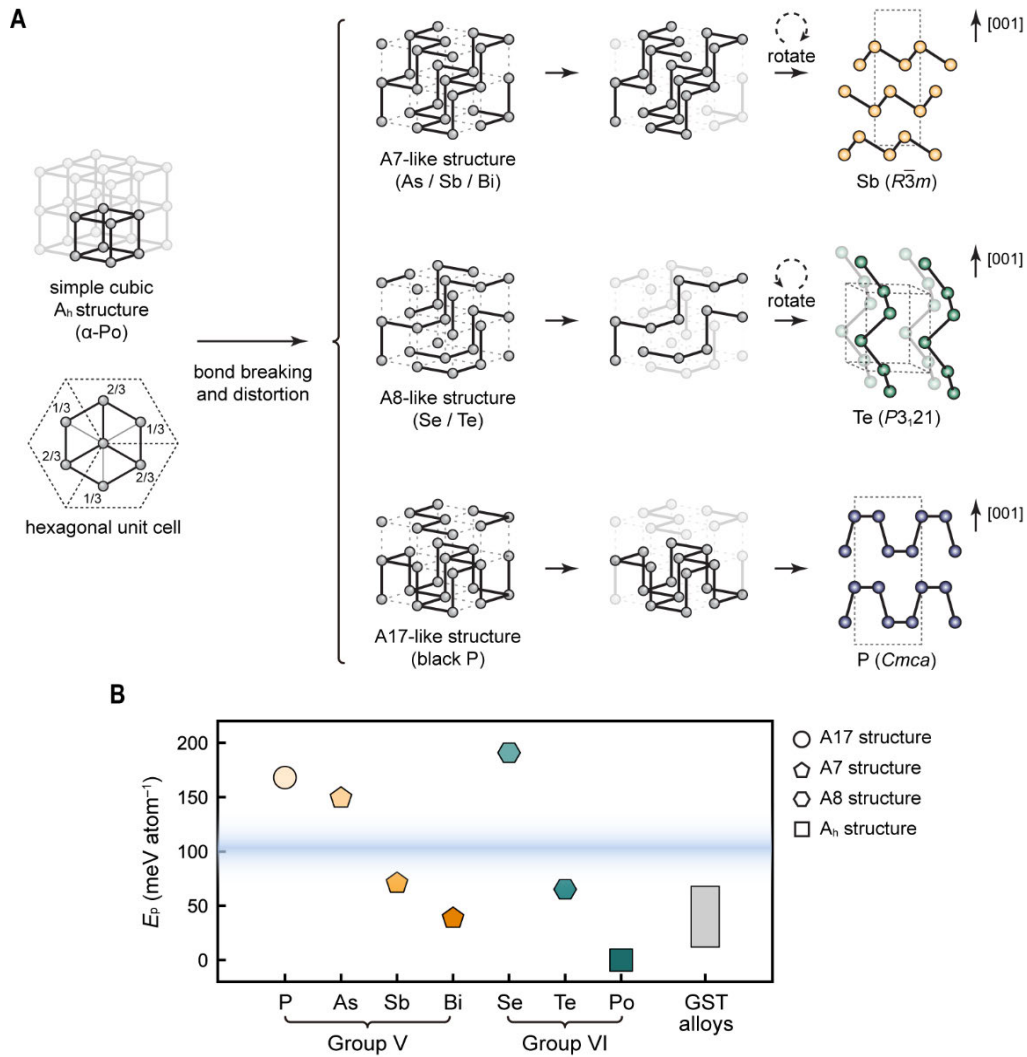

**Fig. S1.** Three-dimensional Peierls distortion and the role of transient cubic-like motifs. (**A**) Transitions from a simple cubic structure to ground-state Sb ($R\bar{3}m$), Te ($P3_121$), and P ($Cmca$) structures via bond breaking and distortion, drawn partly based on Ref. (*S24*) and Ref. (*S25*). (**B**) The calculated cubic-phase formation penalty, $E_p$, defined as the energy difference between the hypothetical cubic phase and the ground-state crystalline phase for various group-V / -VI elements and Ge–Sb–Te (GST) compounds. A gap at $E_p \approx 100$ meV atom$^{-1}$ (indicated by the shaded blue line) clearly separates the lighter, third- and fourth-period elements (i.e., P, As, Se) and the heavier, fifth- and sixth-period elements (i.e., Sb, Bi, Te, Po). Moreover, many Ge–Sb–Te (GST) phase-change memory materials also have $E_p$ values of less than 100 meV atom$^{-1}$, which quickly crystallize into a metastable cubic phase within nanoseconds.

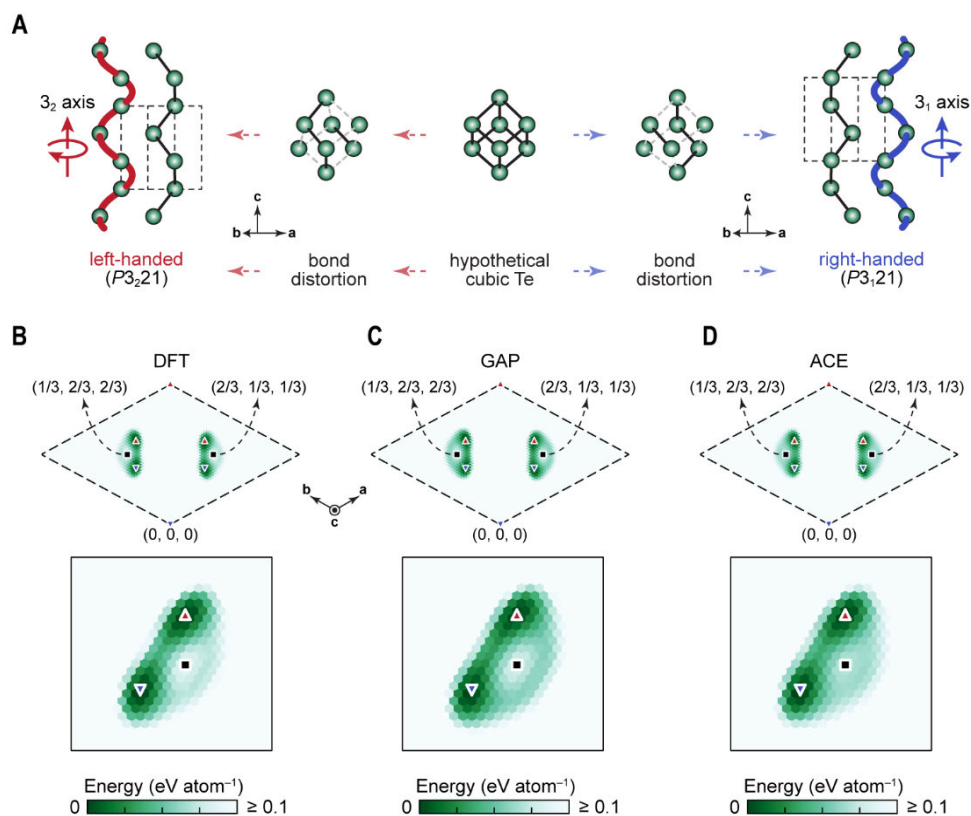

**Fig. S2.** The microscopic route to chirality in crystalline Te. (**A**) Transitions from a simple cubic structure to the ground-state Te ($P3_121$ or $P3_221$) structures via bond breaking and distortion. The chiral helices formed by the left-handed ($3_2$) and right-handed ($3_1$) screw axes are emphasized by thick red and blue lines, respectively, drawn in a style similar to Ref. (*S50*). This panel is reproduced from Fig. 1A in the main text. (**B**–**D**) The energy profile of the projection along the *c*-axis, showing the possible transition between the hypothetical cubic phase (black square) and two energetically equivalent trigonal phases, i.e., left-handed (red triangle) and right-handed (blue inverted triangle) Te. The results computed using (**B**) DFT, (**C**) the GAP potential, and (**D**) the ACE potential are compared, showing good agreement with each other.

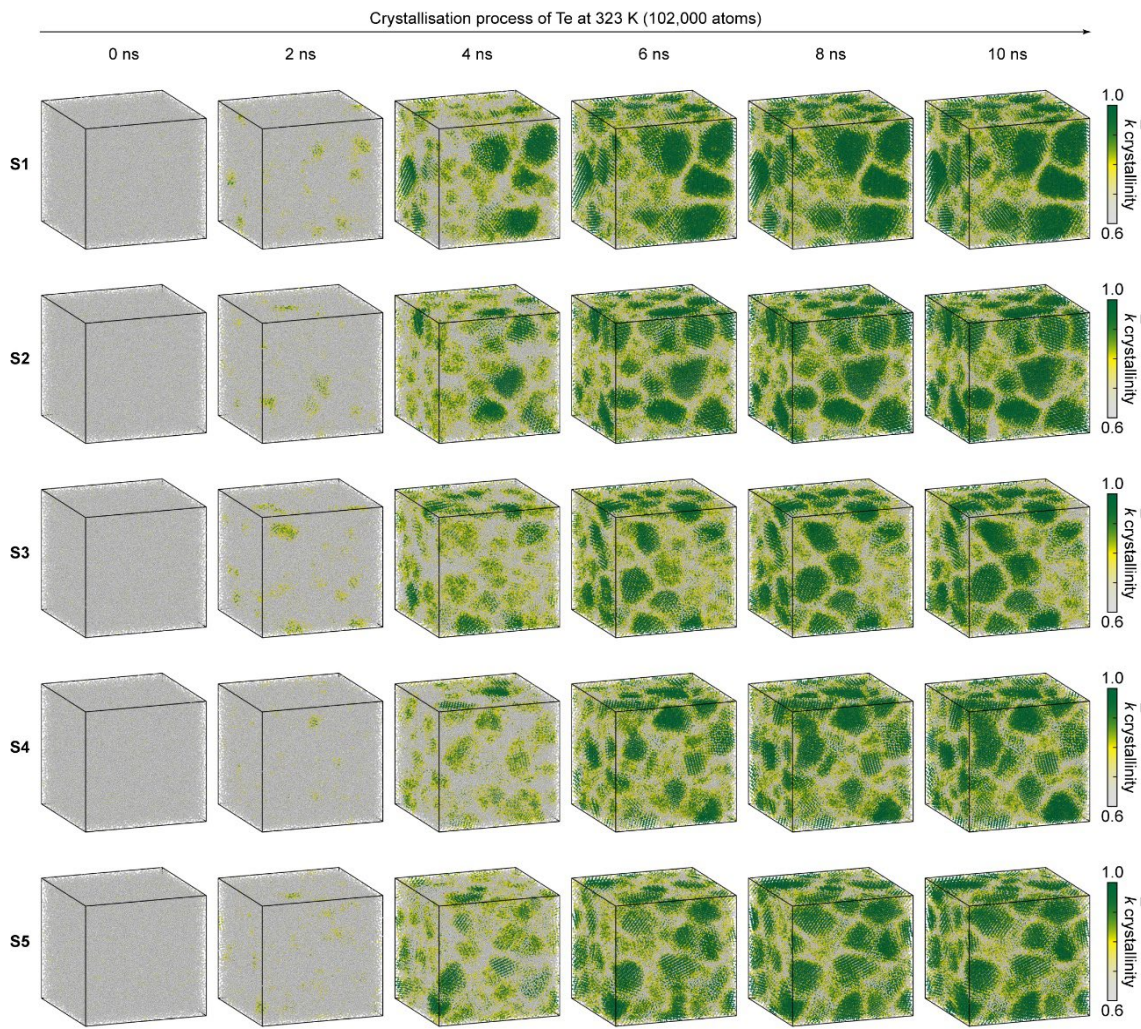

**Fig. S3.** Five independent 10-ns crystallization simulations at 323 K, labeled **S1** to **S5**. Atoms are color-coded using per-atom SOAP crystallinity similarity $\bar{k}$, illustrating the gradual structural ordering from the (disordered) supercooled liquid phase to the crystallized polycrystal. The trajectory of model **S1** is shown in Fig. 2A of the main text.

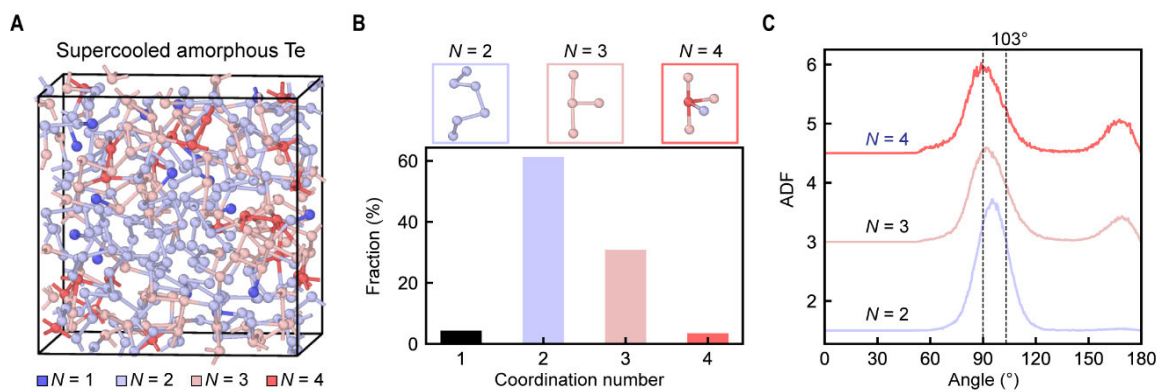

**Fig. S4.** Structural analysis of supercooled amorphous Te at 300 K. (**A**) A snapshot of the supercooled amorphous Te (360-atom) model at 300 K, generated via a melt-quench process using AIMD. Atoms are color coded according to their coordination numbers. (**B**) Distribution of coordination numbers in supercooled amorphous Te. Typical configurations are highlighted and atoms are color coded in the same way as shown in panel (A), expressing the different coordination environments. (**C**) The calculated angular distribution function (ADF), which is split into contributions from atoms with different coordination numbers. An atomic-separation cutoff of 3.2 Å was used in the structural analysis. The two dotted lines highlight the bond angle of 90° (cubic-like motifs) and 103° (chain-like motifs in the crystal).

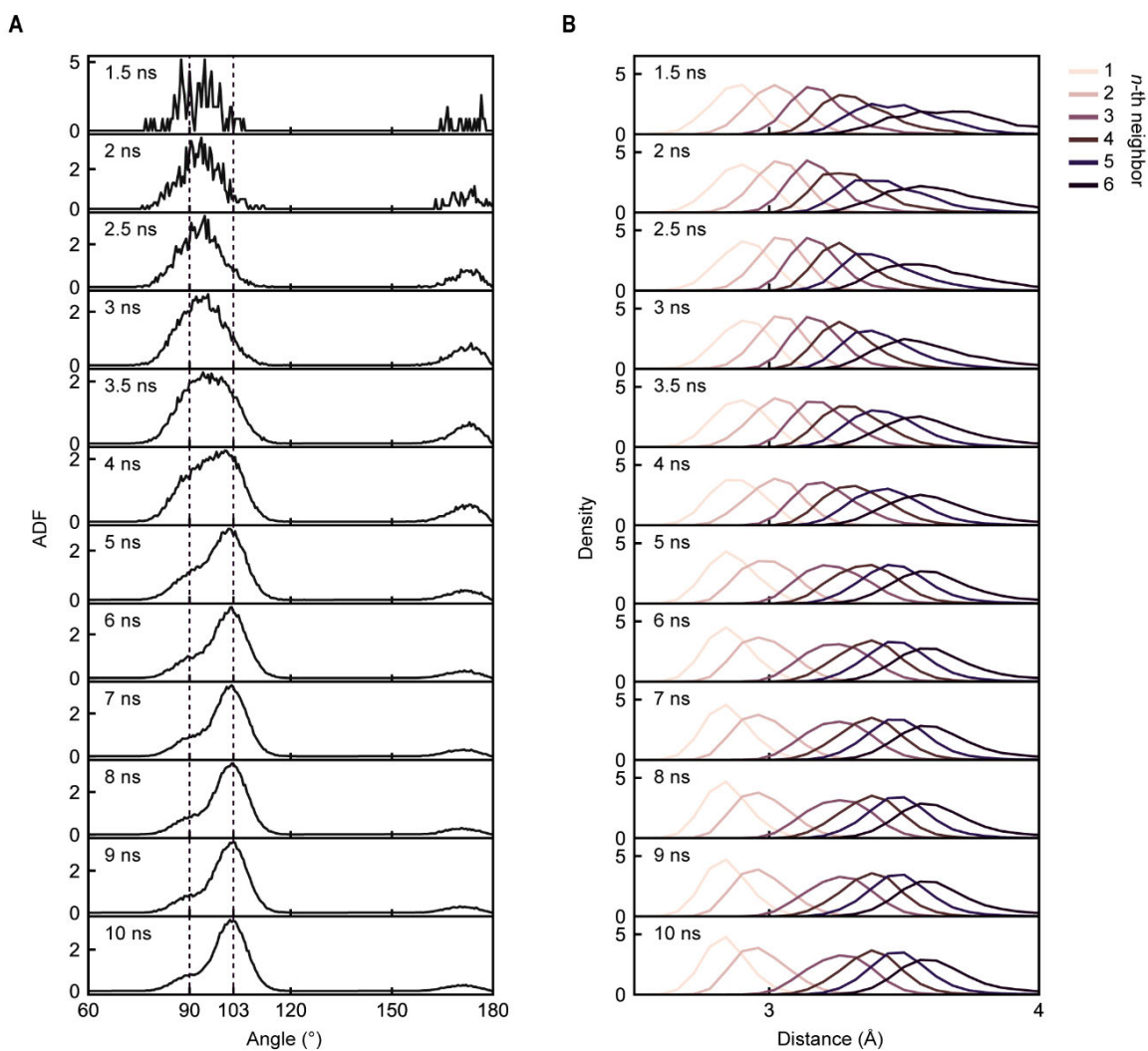

**Fig. S5.** Local structural evolution of crystal-like atoms during the 10-ns crystallization at 323 K. The calculated (**A**) angular distribution function (ADF) and (**B**) *n*-th nearest-neighbor correlation functions show a clear transition from cubic-like motifs (with a coordination number exceeding two and with 90° bond angles) to chain-like helical fragments (with "2 + 4" bonding environments and near-103° bond angles). The two characteristic bond angles, 90° and 103°, were marked by vertical dashed lines in panel (A). In addition, a decrease of the near-180° bond angles, another characteristic of cubic-like motifs, was also observed during the transition.

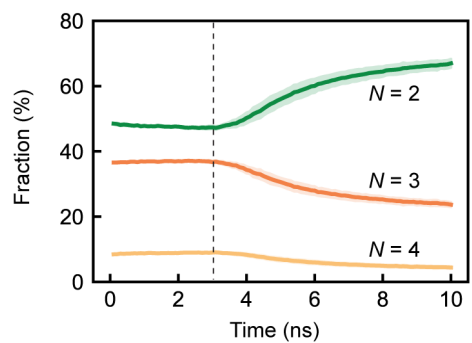

**Fig. S6.** The calculated fraction of atoms having various coordination numbers ($N$ = 2, 3, 4, respectively) during the crystallization process. All curves represent averaged results over 5 independent samples (500 snapshots in total) of ACE-MD simulations, and the shaded areas indicate the standard deviation at each timestep for the 500 snapshots.

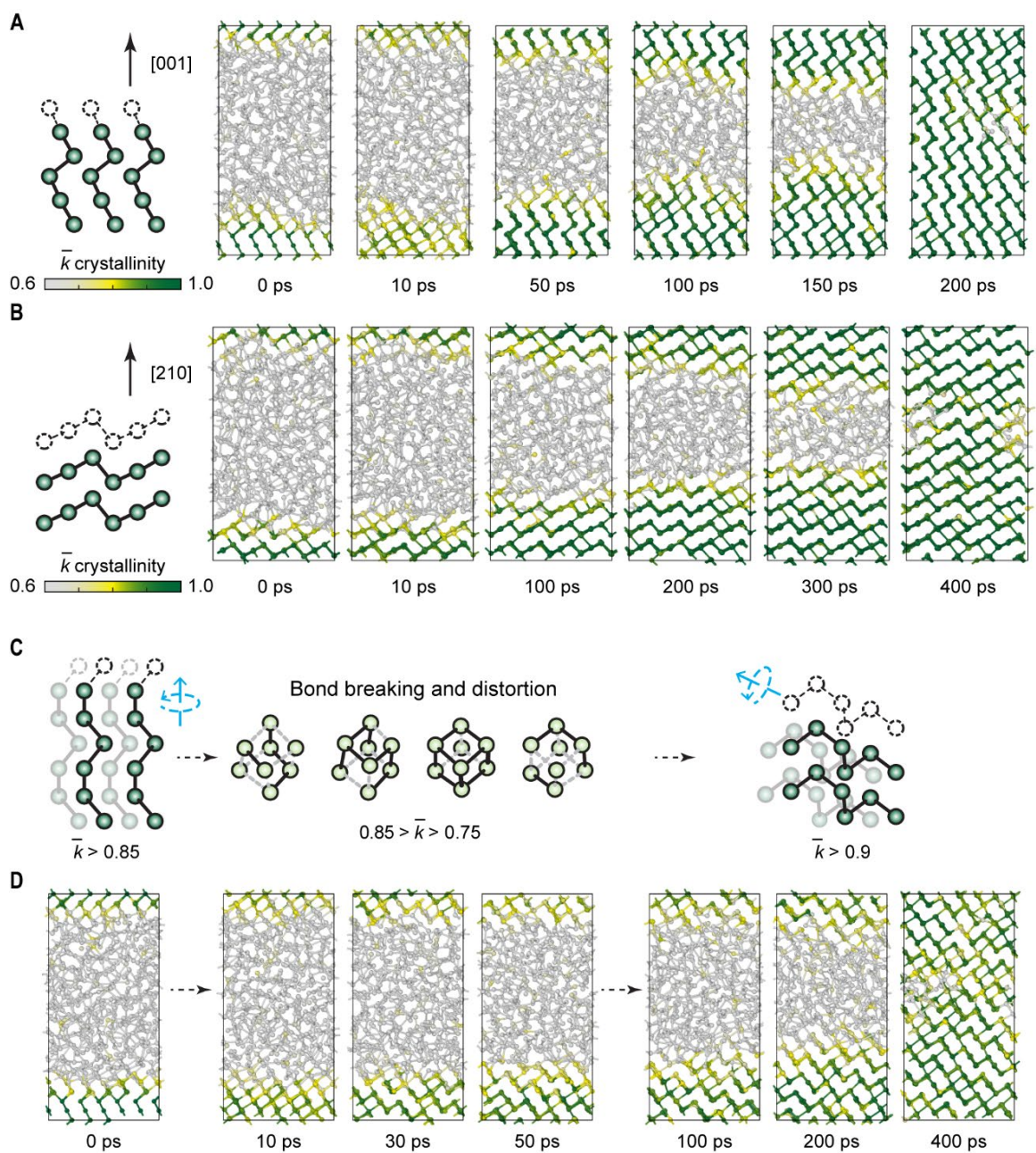

**Fig. S7.** Growth simulations of a templated model (720 atoms) at 423 K using AIMD. (**A**) Growth simulation along the [001] direction, i.e., the chain direction of crystalline Te. (**B**) Growth simulation along the [210] direction, i.e., perpendicular to the chain direction of Te. (**C**) A sketch depicting a change of chain directions via the formation of disordered cubic-like motifs in the growth simulations. (**D**) Growth simulation along the [001] direction; however, the chain direction changed during the growth. Atoms are color-coded using the per-atom SOAP crystallinity similarity $\bar{k}$, illustrating a gradual structural ordering process. We note that the growth process in our AIMD simulations is consistent with the recent AIMD simulations reported in Ref. (*S40*).

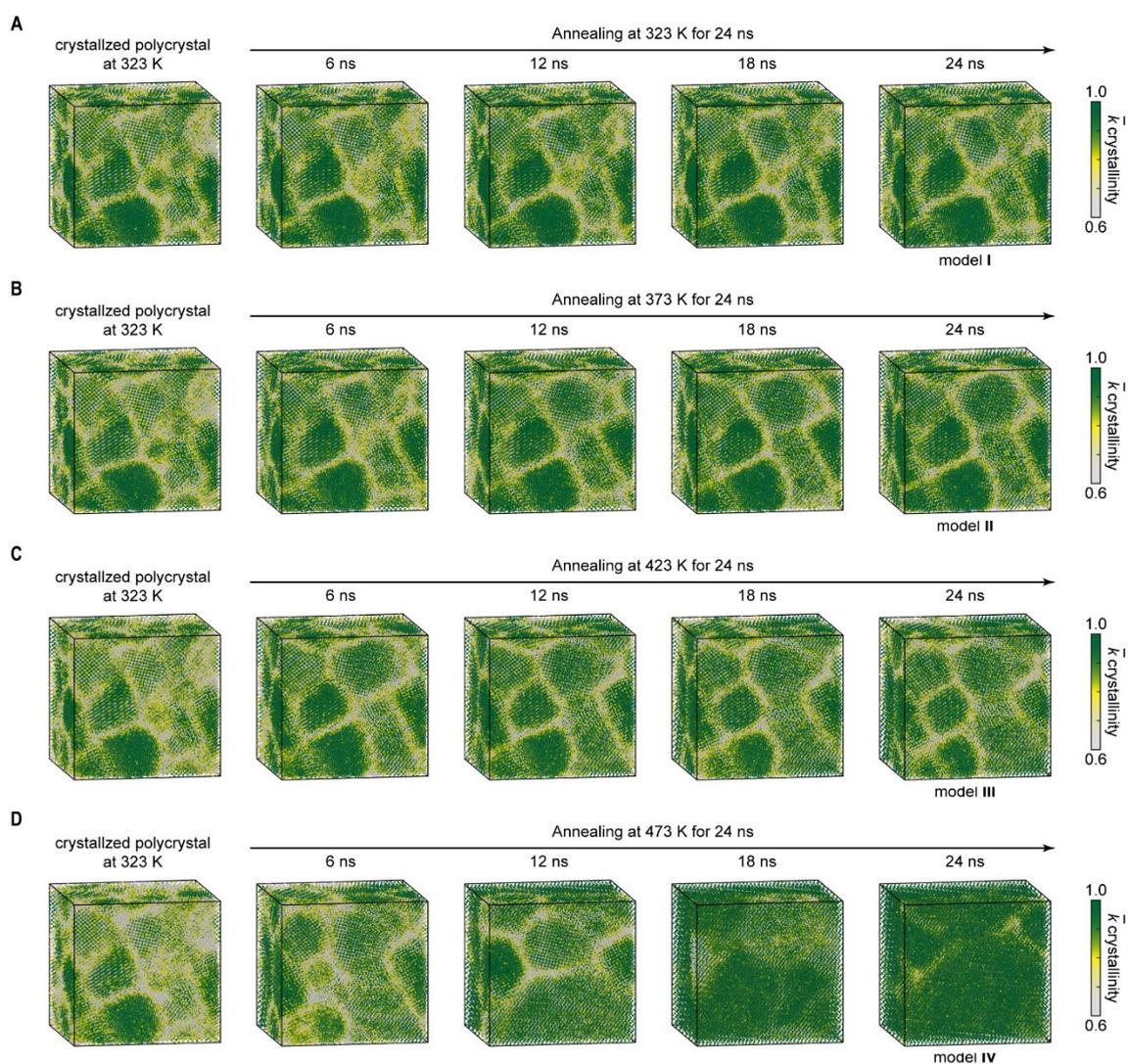

**Fig. S8.** The trajectories of post-annealing simulations. The recrystallized polycrystalline Te sample (as shown in Fig. 2A) was annealed at various temperatures for another 24 ns, including (**A**) 323 K, (**B**) 373 K, (**C**) 423 K, and (**D**) 473 K. Atoms are color-coded using the per-atom SOAP crystallinity similarity $\bar{k}$, illustrating a gradual structural ordering process. The resultant post-annealed structures, labeled as models **I** to **IV**, were used as the starting configurations of the heating simulations in Fig. S14.

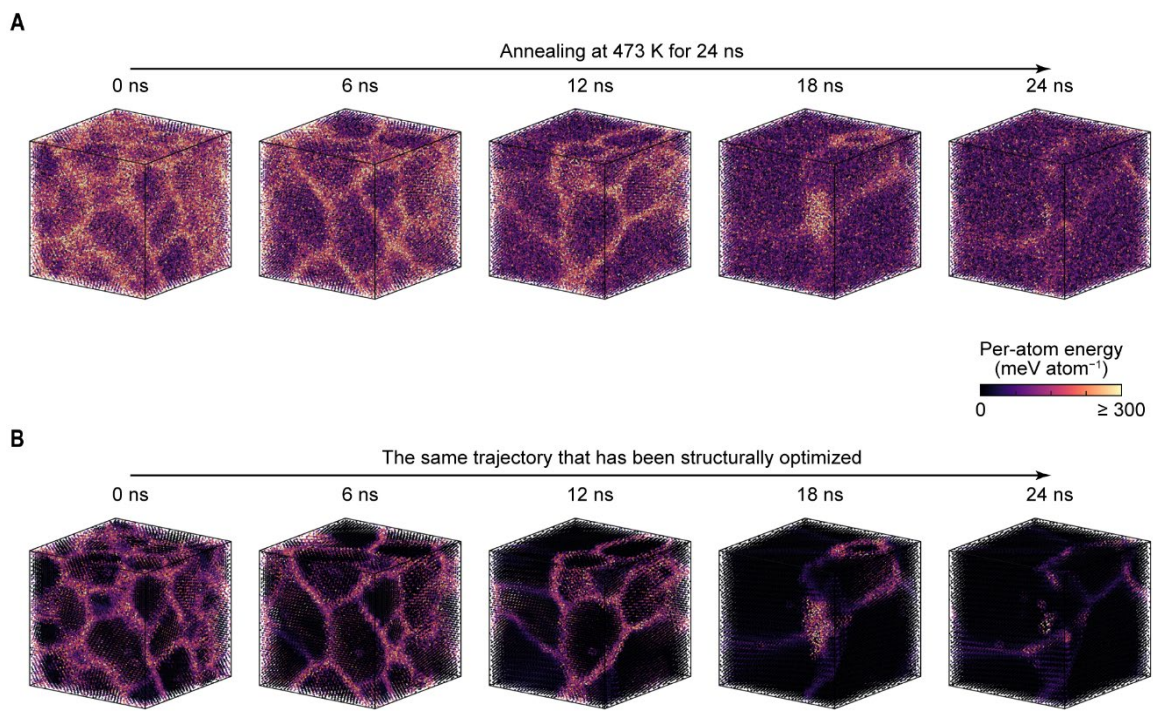

**Fig. S9.** The trajectories of high-temperature annealing simulations at 473 K, with atoms color-coded to indicate the per-atom energy. The structural models were directly taken from the MD trajectories at 473 K: (**A**) without structural optimization; and (**B**) with structural optimization. The reference energy (= 0 meV atom$^{-1}$) is taken from the energy of the optimized α-Te structure.

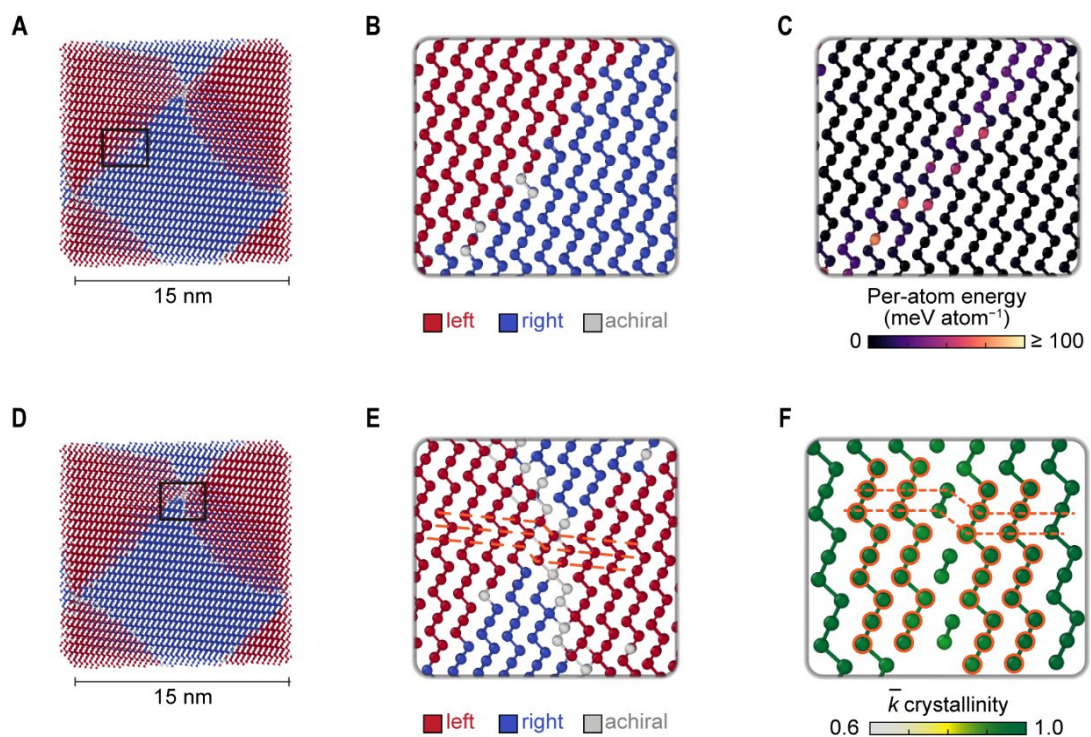

**Fig. S10.** Coexistence of regions with two chiral chains and the formation of screw dislocations in the well-annealed structure. (**A**, **D**) The same slice of the well-annealed structural model. Atoms are color coded using per-atom chirality. The black squares mark two areas where left-handed helices were directly connected with right-handed helices. (**B**, **E**) Magnified view of the regions marked in (A) and (D), respectively. (**C**) The same view as in (B) but with atoms now labeled using the per-atom energy. The reference energy (= 0 meV atom$^{-1}$) is taken from the energy of optimized α-Te. (**F**) Magnified view based on panel (E) where atoms are labeled using the per-atom SOAP crystallinity similarity, $\bar{k}$. Orange dashed lines in (E) and (F) emphasize the trend of forming screw dislocations.

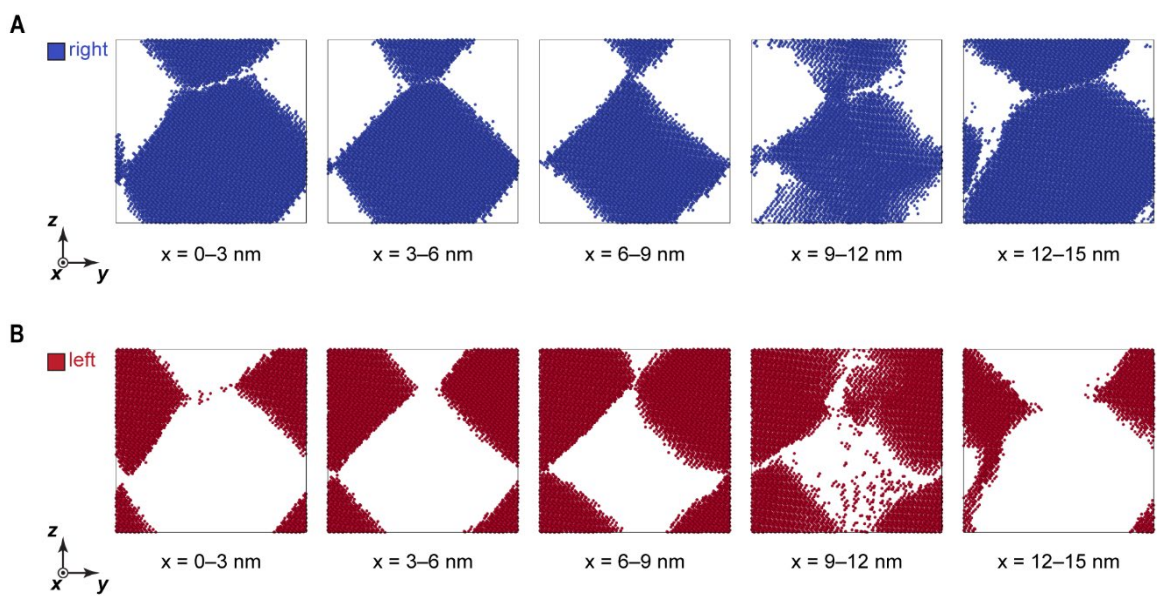

**Fig. S11.** Morphography of two chiral grains in the well-annealed structural model upon high-temperature annealing. (**A**) Snapshots of different parallel slabs (with a thickness of 3 nm) perpendicular to the *x*-axis. Only atoms in right-handed helices are shown. (**B**) Snapshots of different parallel slabs (with a thickness of 3 nm) perpendicular to the *x*-axis. Only atoms in left-handed helices are shown.

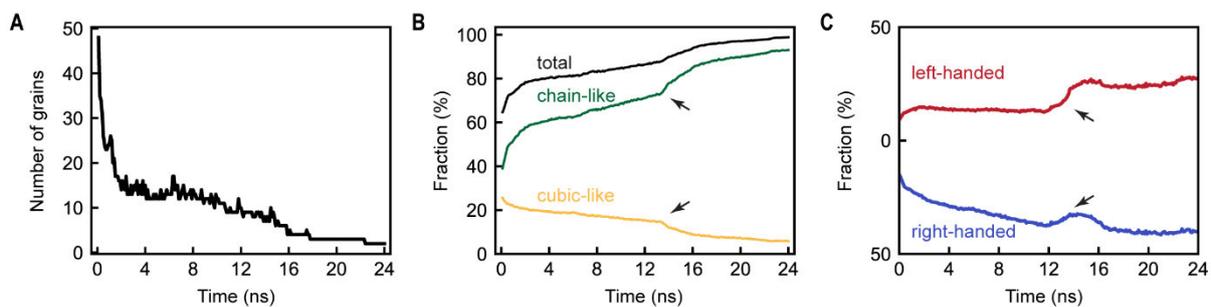

**Fig. S12.** Quantitative structural analysis for high-temperature annealing at 473 K. We show: (**A**) the calculated number of grains; (**B**) the counts of crystal-like (i.e., cubic-like and chain-like) atoms; and (**C**) the calculated fraction of Te atoms in left- and right-handed chiral helices during the annealing process. Black arrows in (B) and (C) indicate multiple nucleation-type chirality transfers upon annealing.

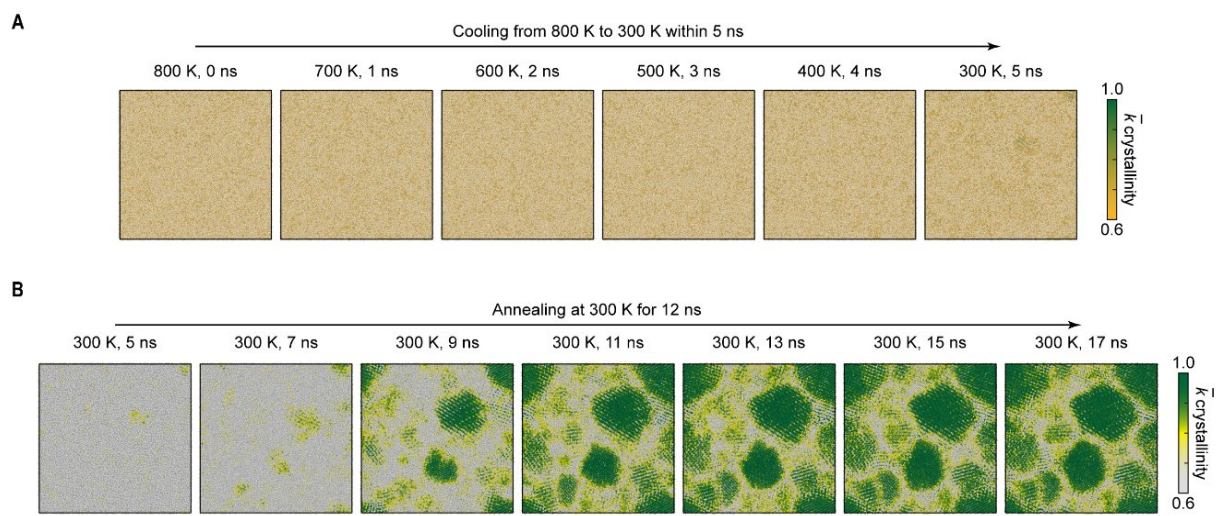

**Fig. S13.** The crystallization of Te during the cooling process. (**A**) The 5-ns cooling process (from 800 to 300 K) of liquid Te. (**B**) The 17-ns crystallization process of Te at 300 K. Atoms are color-coded using two different color schemes to represent the per-atom crystallinity similarity, which separate the cooling process in (A) and the crystallization process (when annealing at 300 K) in (B).

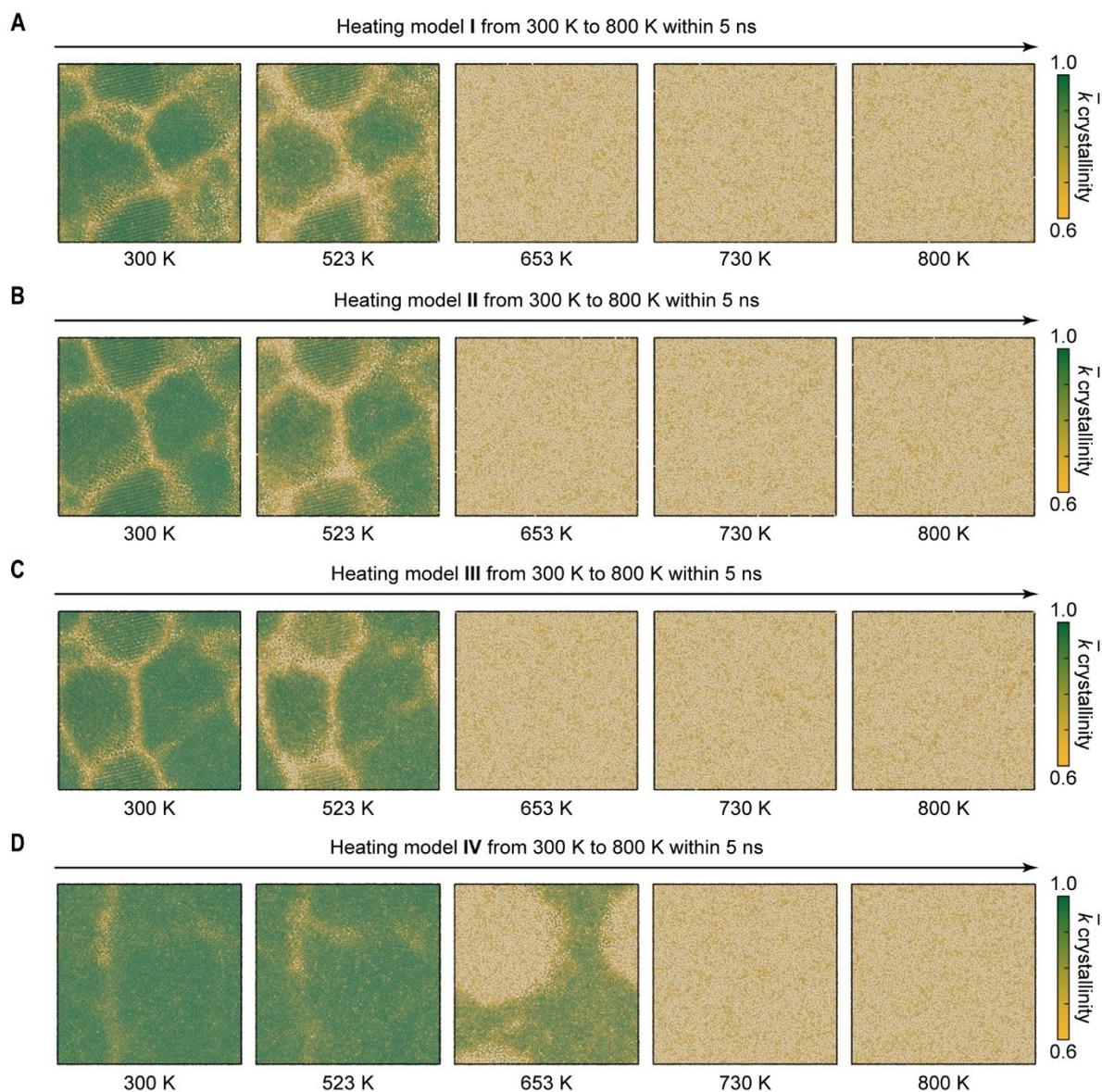

**Fig. S14.** The heating processes of different polycrystalline samples. After the recrystallization of Te (as shown in Fig. 2A) at 323 K, the polycrystalline structure was then post-annealed at four different temperatures, viz. 323, 373, 423, and 473 K, for another 24 ns (cf. Fig. S8). We took the last configuration of these four post-annealing trajectories, labeled as model **I** to **IV**. We then performed four 5-ns heating simulations (from 300 to 800 K) on (**A**) model **I**, (**B**) model **II**, (**C**) model **III**, and (**D**) model **IV**. Atoms are color coded using the per-atom SOAP crystal-similarity $\bar{k}$, illustrating a gradual loss of structural ordering upon heating.

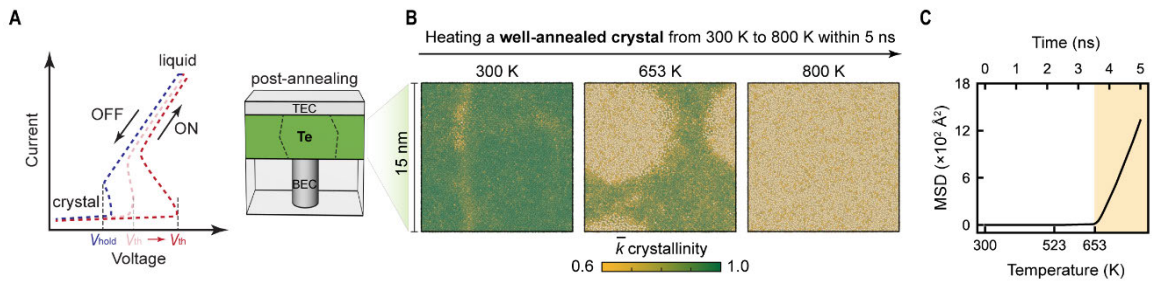

**Fig. S15.** (**A**) A schematic experimental current–voltage curve for Te, demonstrating the increased switch-on voltage upon thermal annealing at high temperatures (*S51*). (**B**) A 5-ns simulated heating process (from 300 to 800 K) of a well-annealed Te crystal. Atoms are color-coded by $\bar{k}$ crystallinity similarity. (**C**) The calculated mean square distance (MSD) of atoms during the heating process, showing a melting at 653 K.

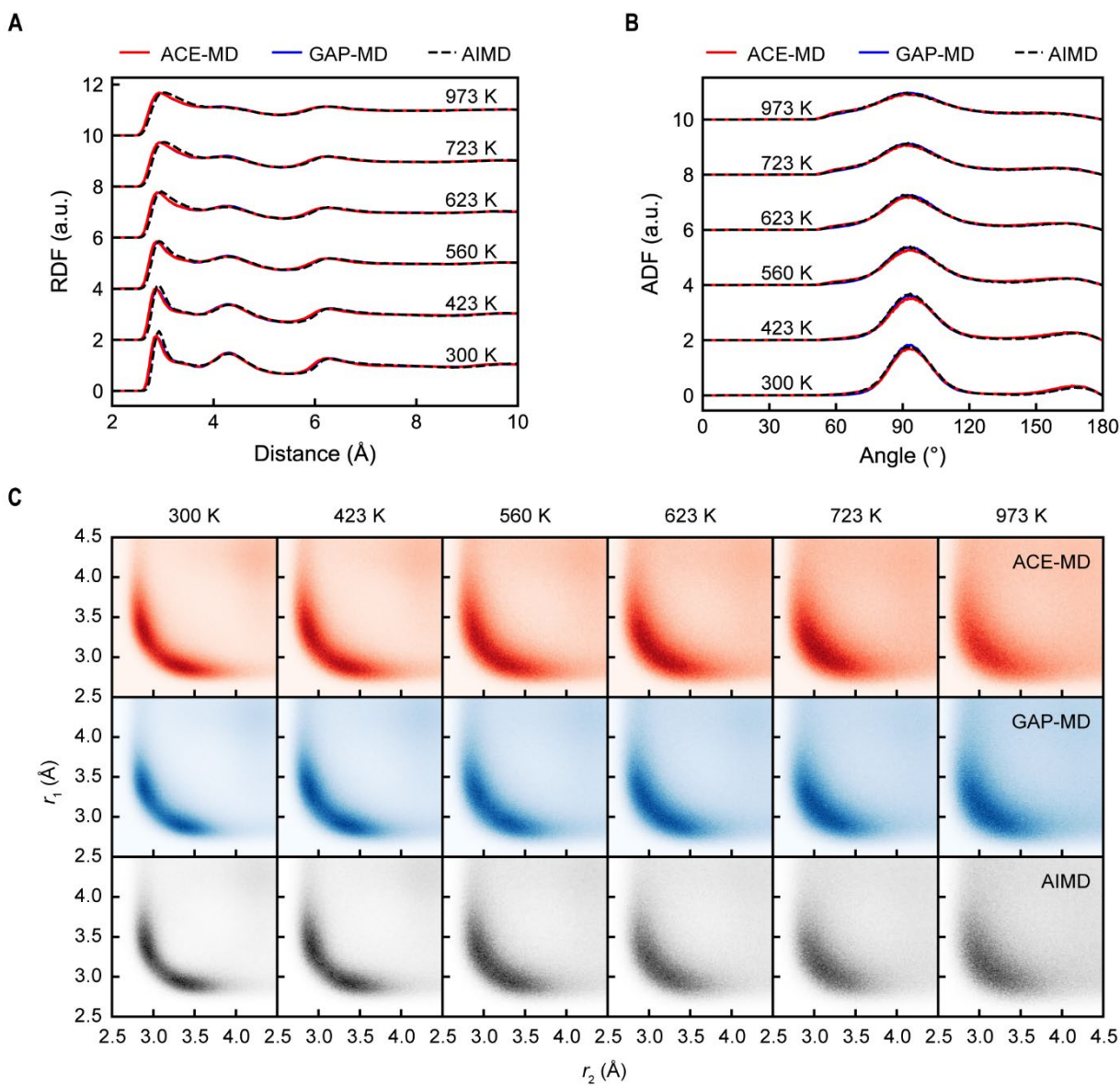

**Fig. S16.** Structural validations for ACE (*red*) and GAP (*blue*) Te potential models as compared to the AIMD reference (*black*). (**A**) Radial distribution function (RDF) computed at various temperatures. (**B**) Angular distribution function (ADF) at different temperatures. (**C**) Angular-limited three-body correlation (ALTBC) functions as indicators for the Peierls distortion in liquid, supercooled liquid and amorphous Te at different temperatures. The ALTBC function expresses the probability of having a bond of length $r_1$ almost aligned with a bond of length $r_2$ (with angular deviations smaller than 30°).

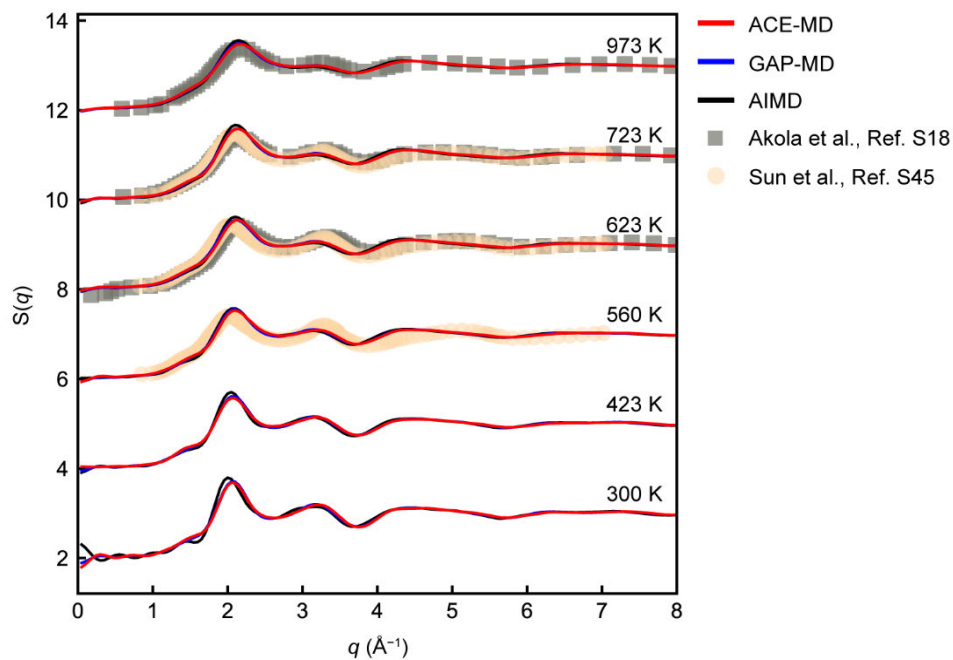

**Fig. S17.** Comparison of structure factors of supercooled liquid and amorphous Te between simulated results (using ACE-MD, GAP-MD, and AIMD) and experimental X-ray diffraction data. Data for different temperatures were calculated, viz. 300, 423, 560, 623, 723, and 973 K. Experimental data were taken from Refs. (*S18*, *S45*).

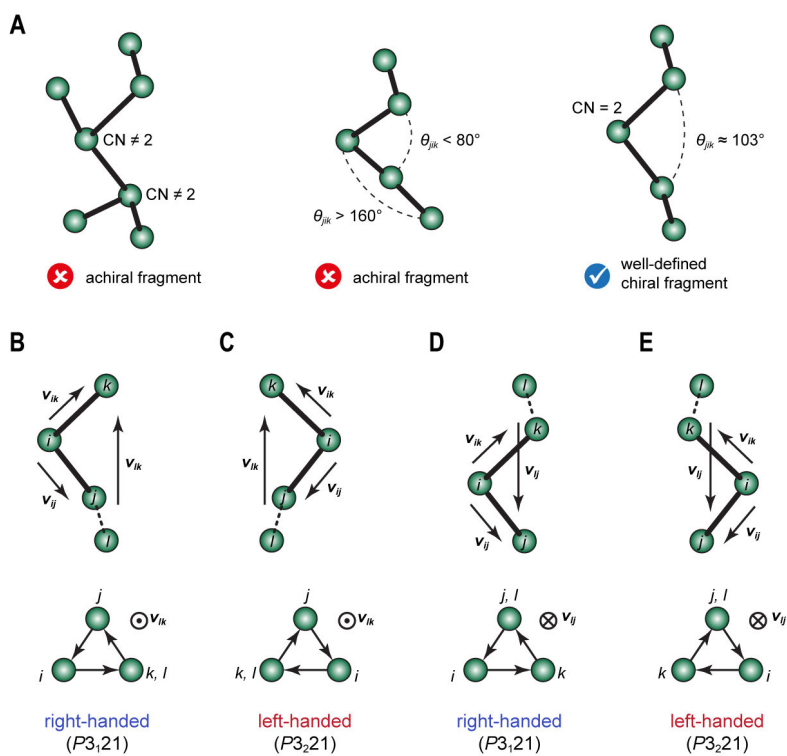

**Fig. S18.** Atomic-level determination of per-atom chirality in Te structures. (**A**) The difference between achiral and well-defined chiral fragments, which can be distinguished based on the computed coordination number (CN) and bond angles ($\theta_{jik}$). (**B**)–(**E**) Four different cases show the atomic-level determination of per-atom chirality based on the geometric connectivity of four bonded atoms in sequence.

**Table S1.** A summary of the reference datasets used for fitting GAP and ACE ML potential models in this work. A single-atom configuration is used as the reference to zero the energies. The "RSS structures" represent the small-scale (fewer than 40 atoms) random structures obtained from GAP-based random structure search (GAP-RSS) processes.

| Type | GAP dataset size | | ACE dataset size | |
|---|---|---|---|---|
| | Cells | Atoms | Cells | Atoms |
| Dimer data | 33 | 66 | 0 | 0 |
| RSS structures | 684 | 7718 | 800 | 9054 |
| Crystalline phases | 284 | 6081 | 304 | 6673 |
| AIMD configurations | 117 | 28788 | 117 | 28788 |
| Standard iterations | 395 | 71100 | 395 | 71100 |
| Domain-specific iterations | 317 | 69840 | 317 | 69840 |
| **Total** | **1830** | **183593** | **1933** | **185455** |

**Table S2.** Computed root-mean-square error (RMSE) values of a test dataset that was not included in the training data of the ML models. The structures in the test dataset can be split into four different categories, including liquid, supercooled liquid, amorphous structures of Te and intermediate configurations of the growth process, which were taken from AIMD simulations.

|  | GAP | | ACE | |
| --- | --- | --- | --- | --- |
|  | RMSE energies (meV atom$^{-1}$) | RMSE forces (meV Å$^{-1}$) | RMSE energies (meV atom$^{-1}$) | RMSE forces (meV Å$^{-1}$) |
| Liquid structures | 13.8 | 164 | 8.1 | 124 |
| Supercooled liquid structures | 41.9 | 169 | 30.4 | 132 |
| Amorphous structures | 52.2 | 164 | 38.3 | 131 |
| Growth configurations | 26.6 | 142 | 16.7 | 107 |

**Supplementary references**

S1. A. P. Bartók, M. C. Payne, R. Kondor, G. Csányi, Gaussian approximation potentials: the accuracy of quantum mechanics, without the electrons. *Phys. Rev. Lett.* **104**, 136403 (2010).

S2. Y. Zhou, W. Zhang, E. Ma, V. L. Deringer, Device-scale atomistic modelling of phase-change memory materials. *Nat. Electron.* **6**, 746–754 (2023).

S3. R. Drautz, Atomic cluster expansion for accurate and transferable interatomic potentials. *Phys. Rev. B* **99**, 014104 (2019).

S4. Y. Lysogorskiy, C. van der Oord, A. Bochkarev, S. Menon, M. Rinaldi, T. Hammerschmidt, M. Mrovec, A. Thompson, G. Csányi, C. Ortner, R. Drautz, Performant implementation of the atomic cluster expansion (PACE) and application to copper and silicon. *npj Comput. Mater.* **7**, 1–12 (2021).

S5. D. F. Thomas du Toit, V. L. Deringer, Cross-platform hyperparameter optimization for machine learning interatomic potentials. *J. Chem. Phys.* **159**, 024803 (2023).

S6. D. F. Thomas du Toit, Y. Zhou, V. L. Deringer, Hyperparameter Optimization for Atomic Cluster Expansion Potentials. arXiv:2408.00656 [Preprint] (2024).

S7. A. Bochkarev, Y. Lysogorskiy, S. Menon, M. Qamar, M. Mrovec, R. Drautz, Efficient parametrization of the atomic cluster expansion. *Phys. Rev. Mater.* **6**, 013804 (2022).

S8. J. Behler, M. Parrinello, Generalized Neural-Network Representation of High-Dimensional Potential-Energy Surfaces. *Phys. Rev. Lett.* **98**, 146401 (2007).

S9. Z. El-Machachi, M. Wilson, V. L. Deringer, Exploring the configurational space of amorphous graphene with machine-learned atomic energies. *Chem. Sci.* **13**, 13720–13731 (2022).

S10. J. D. Morrow, C. Ugwumadu, D. A. Drabold, S. R. Elliott, A. L. Goodwin, V. L. Deringer, Understanding defects in amorphous silicon with million-atom simulations and machine learning. *Angew. Chem. Int. Ed.* **136**, e202403842 (2024).

S11. V. L. Deringer, N. Bernstein, G. Csányi, C. Ben Mahmoud, M. Ceriotti, M. Wilson, D. A. Drabold, S. R. Elliott, Origins of structural and electronic transitions in disordered silicon. *Nature* **589**, 59–64 (2021).

S12. A. P. Thompson, H. M. Aktulga, R. Berger, D. S. Bolintineanu, W. M. Brown, P. S. Crozier, P. J. in 't Veld, A. Kohlmeyer, S. G. Moore, T. D. Nguyen, R. Shan, M. J. Stevens, J. Tranchida, C. Trott, S. J. Plimpton, LAMMPS - a flexible simulation tool for particle-based materials modeling at the atomic, meso, and continuum scales. *Comput. Phys. Commun.* **271**, 108171 (2022).

S13. C. Adenis, V. Langer, O. Lindqvist, Reinvestigation of the structure of tellurium. *Acta Crystallogr., Sect. C: Cryst. Struct. Commun.* **45**, 941–942 (1989).

S14. G. Kresse, J. Furthmüller, Efficient iterative schemes for ab initio total-energy calculations using a plane-wave basis set. *Phys. Rev. B* **54**, 11169–11186 (1996).

S15. G. Kresse, D. Joubert, From ultrasoft pseudopotentials to the projector augmented-wave method. *Phys. Rev. B* **59**, 1758 (1999).

S16. P. E. Blöchl, Projector augmented-wave method. *Phys. Rev. B* **50**, 17953–17979 (1994).

S17. J. Tao, J. P. Perdew, V. N. Staroverov, G. E. Scuseria, Climbing the density functional ladder: nonempirical meta--generalized gradient approximation designed for molecules and solids. *Phys. Rev. Lett.* **91**, 146401 (2003).

S18. J. Akola, R. O. Jones, S. Kohara, T. Usuki, E. Bychkov, Density variations in liquid tellurium: Roles of rings, chains, and cavities. *Phys. Rev. B* **81**, 094202 (2010).

S19. J. Akola, R. O. Jones, Structure and dynamics in amorphous tellurium and Te$_n$ clusters: A density functional study. *Phys. Rev. B* **85**, 134103 (2012).

S20. T. D. Kühne, M. Krack, F. R. Mohamed, M. Parrinello, Efficient and Accurate Car-Parrinello-like Approach to Born-Oppenheimer Molecular Dynamics. *Phys. Rev. Lett.* **98**, 066401 (2007).

S21. T. D. Kühne, M. Iannuzzi, M. Del Ben, V. V. Rybkin, P. Seewald, F. Stein, T. Laino, R. Z. Khaliullin, O. Schutt, F. Schiffmann, D. Golze, J. Wilhelm, S. Chulkov, M. H. Bani-Hashemian, V. Weber, U. Borstnik, M. Taillefumier, A. S. Jakobovits, A. Lazzaro, H. Pabst, T. Muller, R. Schade, M. Guidon, S. Andermatt, N. Holmberg, G. K. Schenter, A. Hehn, A. Bussy, F. Belleflamme, G. Tabacchi, A. Gloss, M. Lass, I. Bethune, C. J. Mundy, C. Plessl, M. Watkins, J. VandeVondele, M. Krack, J. Hutter, CP2K: An electronic structure and molecular dynamics software package - Quickstep: Efficient and accurate electronic structure calculations. *J. Chem. Phys.* **152**, 194103 (2020).

S22. S. Goedecker, M. Teter, J. Hutter, Separable dual-space Gaussian pseudopotentials. *Phys. Rev. B* **54**, 1703 (1996).

S23. A. Stukowski, Visualization and analysis of atomistic simulation data with OVITO–the Open Visualization Tool. *Model. Simul. Mater. Sci. Eng.* **18**, 015012 (2010).

S24. J. K. Burdett, T. J. McLarnan, A study of the arsenic, black phosphorus, and other structures derived from rock salt by bond-breaking processes. I. Structural enumeration. *J. Chem. Phys.* **75**, 5764–5773 (1981).

S25. A. Decker, G. A. Landrum, R. Dronskowski, Structural and electronic Peierls distortions in the elements (A): The crystal structure of tellurium. *Z. Anorg. Allg. Chem.* **628**, 295–302 (2002).

S26. C. H. Champness, R. H. Hoffmann, Conductivity changes associated with the crystallization of amorphous selenium. *J. Non-Cryst. Solids* **4**, 138–148 (1970).

S27. M. Šimečková, A. Hrubý, A study on the crystallization of amorphous arsenic. *Mater. Res. Bull.* **12**, 65–72 (1977).

S28. J. C. Knights, J. E. Mahan, Optical and electrical properties of amorphous arsenic. *Solid State Commun.* **21**, 983–986 (1977).

S29. J. C. Jamieson, Crystal structures adopted by black phosphorus at high pressures. *Science* **139**, 1291–1292 (1963).

S30. H. J. Beister, K. Strössner, K. Syassen, Rhombohedral to simple-cubic phase transition in arsenic under pressure. *Phys. Rev. B* **41**, 5535–5543 (1990).

S31. Y. Akahama, M. Kobayashi, H. Kawamura, Structural studies of pressure-induced phase transitions in selenium up to 150 GPa. *Phys. Rev. B* **47**, 20–26 (1993).

S32. O. Hunderi, Optical properties of crystalline and amorphous bismuth films. *J. Phys. F: Met. Phys.* **5**, 2214 (1975).

S33. M. Salinga, B. Kersting, I. Ronneberger, V. P. Jonnalagadda, X. T. Vu, M. Le Gallo, I. Giannopoulos, O. Cojocaru-Mirédin, R. Mazzarello, A. Sebastian, Monatomic phase change memory. *Nat. Mater.* **17**, 681–685 (2018).

S34. C. Zhao, H. Batiz, B. Yasar, H. Kim, W. Ji, M. C. Scott, D. C. Chrzan, A. Javey, Tellurium single-crystal arrays by low-temperature evaporation and crystallization. *Adv. Mater.* **33**, e2100860 (2021).

S35. J. Hegedus, S. R. Elliott, Computer-simulation design of new phase-change memory materials. *Phys. Status Solidi A* **207**, 510–515 (2010).

S36. X. Shen, Y. Zhou, H. Zhang, V. L. Deringer, R. Mazzarello, W. Zhang, Surface effects on the crystallization kinetics of amorphous antimony. *Nanoscale* **15**, 15259–15267 (2023).

S37. W. Zhang, A. Thiess, P. Zalden, R. Zeller, P. H. Dederichs, J. Y. Raty, M. Wuttig, S. Blügel, R. Mazzarello, Role of vacancies in metal-insulator transitions of crystalline phase-change materials. *Nat. Mater.* **11**, 952–956 (2012).

S38. J. Da Silva, A. Walsh, H. Lee, Insights into the structure of the stable and metastable $(GeTe)_m(Sb_2Te_3)_n$ compounds. *Phys. Rev. B* **78**, 224111 (2008).

S39. T. Matsunaga, Y. Kubota, N. Yamada, Structures of stable and metastable $Ge_2Sb_2Te_5$, an intermetallic compound in the $GeTe$-$Sb_2Te_3$ pseudobinary systems. *Acta Crystallogr., Sect. B* **60**, 685 (2004).

S40. Y. Sun, B. Li, T. Yang, Q. Yang, H. Yu, T. Gotoh, C. Shi, J. Shen, P. Zhou, S. R. Elliott, H. Li, Z. Song, M. Zhu, Nanosecond phase-transition dynamics in elemental tellurium. *Adv. Funct. Mater.*, DOI: 10.1002/adfm.202408725 (2024).

S41. N. Bernstein, G. Csányi, V. L. Deringer, De novo exploration and self-guided learning of potential-energy surfaces. *npj Comput. Mater.* **5**, 1–9 (2019).

S42. V. L. Deringer, C. J. Pickard, G. Csányi, Data-driven learning of total and local energies in elemental boron. *Phys. Rev. Lett.* **120**, 156001 (2018).

S43. A. Jain, S. P. Ong, G. Hautier, W. Chen, W. D. Richards, S. Dacek, S. Cholia, D. Gunter, D. Skinner, G. Ceder, K. A. Persson, Commentary: The Materials Project: A materials genome approach to accelerating materials innovation. *APL Mater.* **1**, 011002 (2013).

S44. J. D. Morrow, J. L. A. Gardner, V. L. Deringer, How to validate machine-learned interatomic potentials. *J. Chem. Phys.* **158**, 121501 (2023).

S45. P. Sun, G. Monaco, P. Zalden, K. Sokolowski-Tinten, J. Antonowicz, R. Sobierajski, Y. Kajihara, A. Q. R. Baron, P. Fuoss, A. C. Chuang, J.-S. Park, J. Almer, J. B. Hastings, Structural changes across thermodynamic maxima in supercooled liquid tellurium: A water-like scenario. *Proc. Natl. Acad. Sci. U. S. A.* **119**, e2202044119 (2022).

S46. A. P. Bartók, R. Kondor, G. Csányi, On representing chemical environments. *Phys. Rev. B* **87**, 184115 (2013).

S47. Y. Xu, Y. Zhou, X. D. Wang, W. Zhang, E. Ma, V. L. Deringer, R. Mazzarello, Unraveling crystallization mechanisms and electronic structure of phase-change materials by large-scale ab initio simulations. *Adv. Mater.* **34**, 2109139 (2022).

S48. K. A. Vladimirovich, G. O. Nikolaevna, "Chirality properties of modeling water in different aqueous systems" in *Chirality from Molecular Electronic States* (IntechOpen, 2018).

S49. T. Bonald, B. Charpentier, A. Galland, A. Hollocou, Hierarchical graph clustering using node pair sampling. arXiv:1806.01664 [Preprint] (2018).

S50. Z. Dong, Y. Ma, Atomic-level handedness determination of chiral crystals using aberration-corrected scanning transmission electron microscopy. *Nat. Commun.* **11**, 1588 (2020).

S51. J. Shen, S. Jia, N. Shi, Q. Ge, T. Gotoh, S. Lv, Q. Liu, R. Dronskowski, S. R. Elliott, Z. Song, M. Zhu, Elemental electrical switch enabling phase segregation-free operation. *Science* **374**, 1390–1394 (2021).